\documentclass[twocolumn]{aastex61}

\hypersetup{backref=true, pagebackref=true, hyperindex=true, breaklinks=true,colorlinks=true,urlcolor=blue, linkcolor=blue,  citecolor=blue,pagecolor=red, bookmarks=true, bookmarksopen=true}

\received{\today}
\revised{---}
\accepted{---}
\submitjournal{ApJ}

\shorttitle{Sarkis et al.: Characterization of K2-18 with Carmenes}
\shortauthors{Sarkis et al.: Characterization of K2-18 with Carmenes}

\usepackage{natbib}
\usepackage{comment}

\label{new commands}

\newcommand{\mpersec}{\ensuremath{\mathrm{ m \, s^{-1}}}}
\newcommand{\kepler}{{\it Kepler}}
\newcommand{\ktwo}{K2-18\,b}
\newcommand{\ktwoc}{K2-18\,c}

\newcommand{\prot}{\ensuremath{P_\mathrm{rot}}}

\begin{document}

\title{The CARMENES search for exoplanets around M dwarfs: \\ A low-mass planet in the temperate zone of the nearby K2-18}

\label{authors}
\correspondingauthor{Paula Sarkis}
\email{sarkis@mpia.de}

\author{Paula Sarkis}
\affil{Max-Planck-Institut f\"{u}r Astronomie, K\"{o}nigstuhl 17, D-69117 Heidelberg, Germany}

\author{Thomas Henning}
\affiliation{Max-Planck-Institut f\"{u}r Astronomie, K\"{o}nigstuhl 17, D-69117 Heidelberg, Germany}
 
\author{Martin K\"{u}rster}
\affiliation{Max-Planck-Institut f\"{u}r Astronomie, K\"{o}nigstuhl 17, D-69117 Heidelberg, Germany}

\author{Trifon Trifonov}
\affiliation{Max-Planck-Institut f\"{u}r Astronomie, K\"{o}nigstuhl 17, D-69117 Heidelberg, Germany}

\author{Mathias Zechmeister}
\affiliation{Institut f\"{u}r Astrophysik, Georg-August-Universit\"{a}t, Friedrich-Hund-Platz 1, 37077 G\"{o}ttingen, Germany}

\author{Lev Tal-Or}
\affiliation{Institut f\"{u}r Astrophysik, Georg-August-Universit\"{a}t, Friedrich-Hund-Platz 1, 37077 G\"{o}ttingen, Germany}

\author{Guillem Anglada-Escud\'{e}}
\affiliation{School of Physics and Astronomy, Queen Mary University of London, E1 4NS London, United Kingdom}
\affiliation{Instituto de Astrof\'{i}sica de Andaluc\'{i}a (CSIC), Glorieta de la Astronom\'{i}a s/n, E-18008 Granada, Spain}

\author{Artie P. Hatzes}
\affiliation{Th\"{u}ringer Landessternwarte, Sternwarte 5, 07778 Tautenburg, Germany}

\author{Marina Lafarga}
\affiliation{Institut de Ci\`{e}ncies de l'Espai (CSIC-IEEC), Campus UAB, c/ de Can Magrans s/n, E-08193 Bellaterra, Barcelona, Spain}

\author{Stefan Dreizler}
\affiliation{Institut f\"{u}r Astrophysik, Georg-August-Universit\"{a}t, Friedrich-Hund-Platz 1, 37077 G\"{o}ttingen, Germany}

\author{Ignasi Ribas}
\affiliation{Institut de Ci\`{e}ncies de l'Espai (CSIC-IEEC), Campus UAB, c/ de Can Magrans s/n, E-08193 Bellaterra, Barcelona, Spain}

\author{Jos\'e A. Caballero}
\affiliation{Centro de Astrobiolog\'{i}a (CSIC-INTA), Campus ESAC, Camino Bajo del Castillo s/n, E-28692 Villanueva de la Ca\~nada, Madrid, Spain}
\affiliation{Landessternwarte, Zentrum f\"{u}r Astronomie der Universt\"{a}t Heidelberg, K\"{o}nigstuhl 12, D-69117 Heidelberg, Germany}

\author{Ansgar Reiners}
\affiliation{Institut f\"{u}r Astrophysik, Georg-August-Universit\"{a}t, Friedrich-Hund-Platz 1, 37077 G\"{o}ttingen, Germany}

\author{Matthias Mallonn}
\affiliation{Leibniz Institute for Astrophysics Potsdam (AIP), An der Sternwarte 16, 14482 Potsdam, Germany}

\author{Juan C. Morales}
\affiliation{Institut de Ci\`{e}ncies de l'Espai (CSIC-IEEC), Campus UAB, c/ de Can Magrans s/n, E-08193 Bellaterra, Barcelona, Spain}

\author{Adrian Kaminski}
\affiliation{Landessternwarte, Zentrum f\"{u}r Astronomie der Universt\"{a}t Heidelberg, K\"{o}nigstuhl 12, D-69117 Heidelberg, Germany}

\author{Jes\'us Aceituno}
\affiliation{Centro Astron\'omico Hispano Alem\'an de Calar Alto (CSIC-MPG), C/ Jes\'us Durb\'an Rem\'on 2–2, 4004 Almer\'ia, Spain}

\author{Pedro J. Amado}
\affiliation{Instituto de Astrof\'{i}sica de Andaluc\'{i}a (CSIC), Glorieta de la Astronom\'{i}a s/n, E-18008 Granada, Spain}

\author{Victor J. S. B\'ejar}
\affiliation{Instituto de Astrof\'isica de Canarias, Vía L\'actea s/n, 38205 La Laguna, Tenerife, Spain, and Departamento de Astrof\'isica, Universidad de La Laguna, 38206 La Laguna, Tenerife, Spain}

\author{Hans-J\"urgen Hagen}
\affiliation{Hamburger Sternwarte, Gojenbergsweg 112, D-21029 Hamburg, Germany}
 
\author{Sandra Jeffers}
\affiliation{Institut f\"{u}r Astrophysik, Georg-August-Universit\"{a}t, Friedrich-Hund-Platz 1, 37077 G\"{o}ttingen, Germany}

\author{Andreas Quirrenbach}
\affiliation{Landessternwarte, Zentrum f\"{u}r Astronomie der Universt\"{a}t Heidelberg, K\"{o}nigstuhl 12, D-69117 Heidelberg, Germany}

\author{Ralf Launhardt}
\affiliation{Max-Planck-Institut f\"{u}r Astronomie, K\"{o}nigstuhl 17, D-69117 Heidelberg, Germany}

\author{Christopher Marvin}
\affiliation{Institut f\"{u}r Astrophysik, Georg-August-Universit\"{a}t, Friedrich-Hund-Platz 1, 37077 G\"{o}ttingen, Germany}

\author{David Montes}
\affiliation{Departamento de Astrof\'isica y Ciencias de la Atm\'osfera, Facultad de Ciencias F\'isicas, Universidad Complutense de Madrid, E-28040 Madrid, Spain}



\begin{abstract} \label{abstract}

K2-18 is a nearby M2.5 dwarf, located at 34 pc
and hosting a transiting planet
which was first discovered by the {\it K2} mission and later confirmed
with {\it Spitzer Space Telescope} observations.
With a radius of $\sim 2 \, R_{\oplus}$
and an orbital period of $\sim 33$ days,
the planet lies in the temperate zone of its host star 
and receives stellar irradiation similar to Earth.
Here we perform radial velocity follow-up observations with 
the visual channel of CARMENES
with the goal of determining the mass
and density of the planet.
We measure a planetary 
semi-amplitude of $K_b \sim 3.5$ \mpersec\
and a mass of 
$M_b \sim 9 \, M_{\oplus}$,
yielding a bulk density around 
$\rho_b \sim 4 \, \mathrm{g \,cm^{-3}}$.
This indicates a low-mass planet with a composition
consistent with a solid core and a volatile-rich envelope.
A signal at 9 days was recently reported 
using radial velocity measurements 
taken with the HARPS spectrograph. 
This was interpreted as being due to a second planet. 
We see a weaker,
time and wavelength dependent
signal in the CARMENES data set
and thus favor stellar activity for its origin.
\ktwo\ joins the growing group of low-mass planets 
detected in the temperate zone of M dwarfs.
The brightness of the host star in the near-infrared
makes the system a good target for 
detailed atmospheric studies with the {\it James Webb Space Telescope}.

\end{abstract}





\section{Introduction} \label{sec:intro}

The search for exoplanets around M dwarfs
has expanded steadily over the last years because 
it allows the first detections of
low-mass planets in their habitable zones.
Because of their low masses and small radii, 
compared to Sun-like stars, 
relatively large radial velocity (RV) amplitudes 
and transit depths can occur.
Moreover, the low luminosity of M dwarfs implies that the 
planets in the habitable zones of these stars
are located closer to the star
and at shorter orbital periods.
Indeed, the recent discoveries of Earth-like 
low-mass planets orbiting in the habitable zones of M stars 
have demonstrated the importance of these targets 
\citep[e.g.,][]{Crossfield:2015,Gillon:2017, Dittmann:2017, Bonfils:2017},
with perhaps the most exciting discovery
being the detection of a potentially habitable planet 
orbiting our stellar neighbor Proxima Centauri
\citep{Anglada-Escude:2016}.

However, a major challenge in detecting low-mass planets
around M dwarfs 
is the activity of their host stars.
Common features of activity 
are dark starspots and bright plage regions, 
both of which can break the flux balance 
between the blueshifted approaching hemipshere
and the redshifted receding hemisphere. 
As a result, active regions may produces distortions 
in the spectral lines
that give rise to RV variations.
Such activity signals could obscure or hinder
the detection of low-mass planets 
or even mimic the presence of a false planetary signal.
They often appear at the stellar rotation period
and its harmonics \citep{Boisse:2011}.
For example,
\cite{Robertson:2014} and \cite{Hatzes:2016} showed that 
the RV variations associated to GJ 581d
correlate with the H$\alpha$ index,
which is a magnetic activity indicator. 
This is an indication that GJ 581d is most likely not a planet 
and its RV signal is a harmonic of the stellar rotation period
(but see \cite{Anglada-Escude:2015}).

There are several on-going and future precise RV surveys
whose main goal is to search for terrestrial planets
around M dwarfs, including CARMENES \citep{Quirrenbach:2014},
HPF \citep{Mahadevan:2012},
IRD \citep{Tamura:2012}, 
NIRPS \citep{Bouchy:2017},
and SPIRou \citep{Artigau:2014}.
Stellar activity poses a challenge in finding these planets. 
It is even more difficult to 
disentangle the planetary signal from the activity signal
when the orbital period of the planet is close to 
that of the stellar activity.
The stellar rotation periods of early M dwarfs often coincide
with the orbital periods 
of planets in their habitable zones \citep{Newton:2016}.
Therefore, correcting for stellar activity 
requires the rotational period
to be accurately known. 
Contemporaneous photometry 
is thus crucial to determine the 
rotational period and to 
differentiate between planetary and 
activity signals.
Another powerful way
is to obtain RV measurements at different wavelengths. 
This enables the comparison between the blue part and 
the red part of the spectrum, 
where, unlike a wavelength-independent Keplerian signal, 
RV signals due to activity are wavelength dependent \citep{Reiners:2010}.

In this work, we aim to estimate the mass
and hence the density of the transiting planet \ktwo\ 
by analyzing the RV signals obtained with CARMENES.
The host star is a nearby M2.5\,V star.
\ktwo\ receives approximately the same level of
stellar irradiation as Earth
and orbits in the temperate zone,
where water could exist in its liquid form.
Two planetary transits were observed with \kepler\
as part of the {\it K2} mission during Campaign 1
\citep{Montet:2015}.
Later, \cite{Benneke:2017} confirmed the planetary nature 
of the transit signal by 
observing the same transit depth 
at a different wavelength, 4.5 $\mu$m,
with the {\it Spitzer Space Telescope}.
These observations validated 
the signal seen in the {\it K2} photometry 
and ruled out the alternative scenario of 
two long-period planets with similar sizes,
each transiting once during the {\it K2} observations.

\cite{Cloutier:2017} presented
precise RV follow-up observations of K2-18
performed with the HARPS spectrograph \citep{Mayor:2003}.
They estimated the mass and density of \ktwo,
and additionally reported the discovery 
of a second non-transiting planet in the system. 
In this paper we first present the results
of independent RV observations and analysis 
of the system.
Second, we compare the results of both 
CARMENES and HARPS campaigns, 
and finally combine both data sets
to refine the parameters of the system.

For this study, observations were carried out with
the high-resolution spectrograph CARMENES \citep{Quirrenbach:2014},
which is the first operational spectrograph 
that is designed to 
obtain precise RVs
in the visible and in the near-infrared (NIR) simultaneously.
Its design was motivated by the scientific goal
of detecting low-mass planets in the habitable zone of 324 M dwarfs
\citep{Reiners:2017}.
\cite{Trifonov:2018} demonstrated that CARMENES
is indeed capable of discovering rocky planets around low-mass stars.
\cite{Reiners:2018} reported the discovery 
of the first CARMENES exoplanet 
from the survey 
around HD147379\,b, an M0.0V star.
We also acquired simultaneous photometric observations
in the Johnson {\it B} and Cousins {\it R} filters
to estimate the stellar rotation period.

As the optimization of the NIR channel 
to the precision required to carry out such studies is still ongoing, 
we concentrate on the data taken in the visual channel (VIS),
which contains several activity indicators 
and covers redder orders than HARPS.
Where appropriate, 
we will address the data obtained by the visual channel as CARMENES-VIS
and address the instrument as a whole as CARMENES.

The paper is structured as follows:
In Section~\ref{sec:data} we present the 
spectroscopic and photometric data sets.
In Section~\ref{sec:activity} 
we estimate the stellar rotation period 
and analyse the stellar activity. 
Section~\ref{sec:rv}
describes different tests that we performed
to analyse the RV data set and 
compare our results with the results of \cite{Cloutier:2017}.
In Section~\ref{sec:both-sets} we refine 
the planetary parameter by combining 
both CARMENES and HARPS data sets. 
In Section \ref{sec:results} we discuss our results,
and give our conclusions in Section \ref{sec:conclusion}.


\section{Data} \label{sec:data}

\subsection{Radial Velocities}

CARMENES
(Calar Alto 
high-Resolution search for 
M dwarfs with 
Exo-earths with 
Near-infrared and optical 
Echelle Spectrographs)
is a pair of high-resolution echelle spectrographs
\citep{Quirrenbach:2014}
mounted on the 3.5 m telescope 
of the Calar Alto Observatory (CAHA) in Spain.
The VIS channel
covers the wavelength 
range from 0.52 to 0.96 $\mu$m and has a
spectral resolution $R = 94,600$ \citep{Quirrenbach:2016},
with a demonstrated precision similar to HARPS
and better than Keck/HIRES
\citep{Trifonov:2018}.

We monitored K2-18 between December 2016 
and June 2017 with CARMENES.
In total 58 spectra were obtained
which were reduced and extracted 
using the CARACAL pipeline \citep{Zechmeister:2018,Caballero:2016}.
The pipeline implements the standard method for reducing a spectrum,
i.e. each spectrum was corrected for bias, flatfield, and cosmic rays,
followed by a flat-relative optimal extraction of the 
1D spectra \citep{Zechmeister:2014} and wavelength calibration.
In order to get precise RVs, 
we use the data products from the SERVAL pipeline
\citep{Zechmeister:2018},
which uses a least-squares fitting algorithm.
Following the approach by \cite{Anglada-Escude:2012},
a high signal-to-noise ratio spectrum is constructed 
by a suitable combination of the observed spectra 
and used as a template to measure the RVs.
The SERVAL-estimated RVs were additionally corrected 
for small night-to-night systematic zero-point variations,
as explained in \cite{Trifonov:2018}.
The origin of the offsets is still unclear
but they are probably due to systematics in the wavelength solution
and a slow drift in the calibration source during the night.
The time series is shown in the left panel of Figure~\ref{fig:rvplot}.
The optical differential RV measurements
and the activity indicators (see Section \ref{sec:activity}) 
used in the analysis
are reported in Table \ref{tab:data-carmenes}.

\subsection{Photometry}

We monitored the host star K2-18 for photometric variability 
with the robotic 1.2\,m twin-telescope STELLA on Tenerife \citep{Strassmeier:2004}
and its wide-field imager WiFSIP. 
From February 2017 until June 2017, 
we obtained blocks of four exposures in Johnson {\it B}
and four exposures in Cousins {\it R} over 33 nights. 
The exposure time was 120 seconds in {\it B} and 60 seconds in {\it R}. 
The data reduction was performed identically to previous host star monitoring campaigns 
with STELLA \citep{Mallonn:2015, Mallonn:2016}. 
The bias and flatfield correction was made with the STELLA data reduction pipeline. 
We performed aperture photometry with the software Source Extractor 
\citep{Bertin:1996}. 
For differential photometry 
we divided the flux of the target 
by the combined flux of an ensemble of comparison stars. 
The flux of these stars was combined after giving them an optimal weight 
according to the scatter in their light curves \citep{Broeg:2005}. 
We verified that the selection of comparison stars did not significantly affect 
the variability signal seen in the differential light curve of K2-18.
The nightly observations were averaged 
and a few science frames were discarded 
due to technical problems. 
The final light curves contain 29 data points 
in {\it B} and 28 data points in {\it R} 
and are shown in Figure~\ref{fig:phot_time_series}.

\section{Rotation Period and Stellar Activity} \label{sec:activity}

The presence of active regions on the surface of a star 
can produce RV variations and hence mimic the presence of a planet
\citep{Robertson:2014, Robertson:2015, Hatzes:2016}. 
A common way to distinguish whether the RV signal
is due to a planet or due to activity
is to check for periodicities in the 
activity indicators
and for photometric variability.
We present first the analysis
of the stellar photometric variability 
(Section \ref{sec:activity-phot}),
then the analysis of the spectroscopic activity indicators 
(Section \ref{sec:activity-spectro}),
and finally compare the
chromospheric and photospheric variability (Section \ref{sec:chromoVSphoto}).

\subsection{Photometric Variability} \label{sec:activity-phot}

Active regions, 
in the form of dark spots and bright plages,
rotate with the stellar surface
and produce photometric as well as RV variability.
The observed RV signal is often 
detected at the stellar rotation period (\prot)
and its harmonics (\prot/2, \prot/3, ...)
\citep{Boisse:2011}.
Its amplitude and phase may also vary in time 
due to the evolution of the active regions.
Therefore, contemporaneous 
photometry and RV observations 
are important to determine 
the stellar rotation period 
and to differentiate 
between a planetary and stellar activity signals.

The photometric and spectroscopic observations were performed 
during the same observational season in 2017. 
In order to estimate the stellar rotation period, we 
followed the classical approach by applying the 
Generalized Lomb-Scargle periodogram
\citep[GLS;][]{Zechmeister:2009}
to the photometric data sets.
The GLS analysis showed a peak at $\sim$ 40 days in the {\it B} band 
and a peak at $\sim$ 39 days in the {\it R} band. 
To assess the false alarm probability (FAP) of the signals,
we applied the bootstrap randomization technique
\citep{Bieber:1990,Kuerster:1997}. 
This is done by computing the GLS 
of a set obtained by randomly shuffling the 
observed magnitudes with the times of observations. 
We repeated this 10,000 times 
and the FAP is defined as the number 
of times where the periodogram
of the randomized data sets
shows a GLS power as high as or higher than 
that of the original data set. 
We found that the FAP is $< 10^{-4}$ in the {\it B} band 
and FAP = $2 \times 10^{-4}$ in the {\it R} band.
The upper panel in Figure~\ref{fig:phot_periodogram}
shows the periodogram of the data taken with the {\it B} filter
and the lower panel of those taken with the {\it R} filter.

\begin{figure}[t!]
	\includegraphics[width=0.5\textwidth]{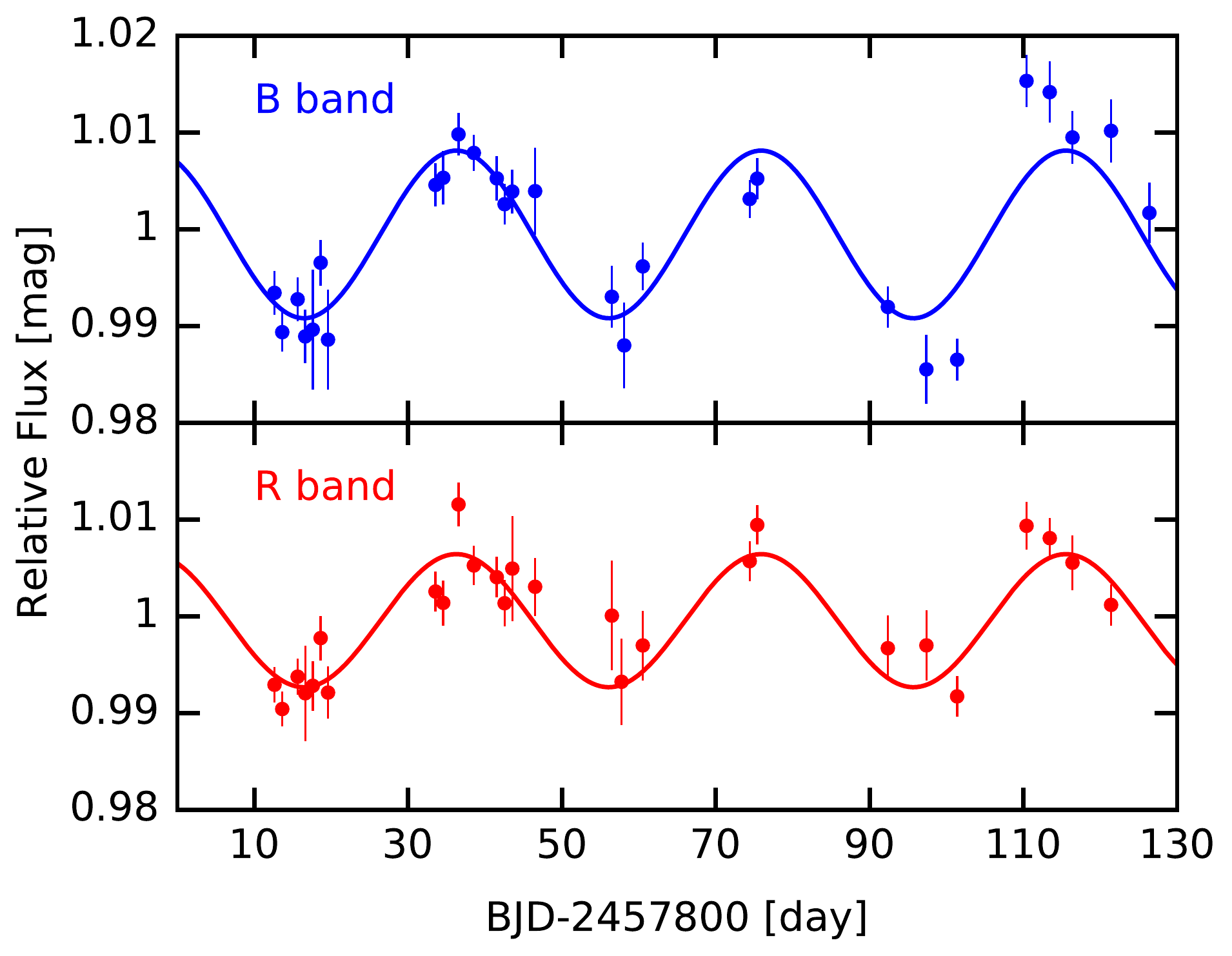}
	\caption{WiFSIP/STELLAR differential photometry of K2-18 taken in {\it B} (upper panel) and in {\it R} (lower panel). The solid curves show the best sine fit to the data. The star shows photometric variations with a semi-amplitude of 0.86\% in the {\it B} band and 0.69\% in the {\it R} band. 
	\label{fig:phot_time_series}}
\end{figure}

\begin{figure}
	\includegraphics[width=0.5\textwidth]{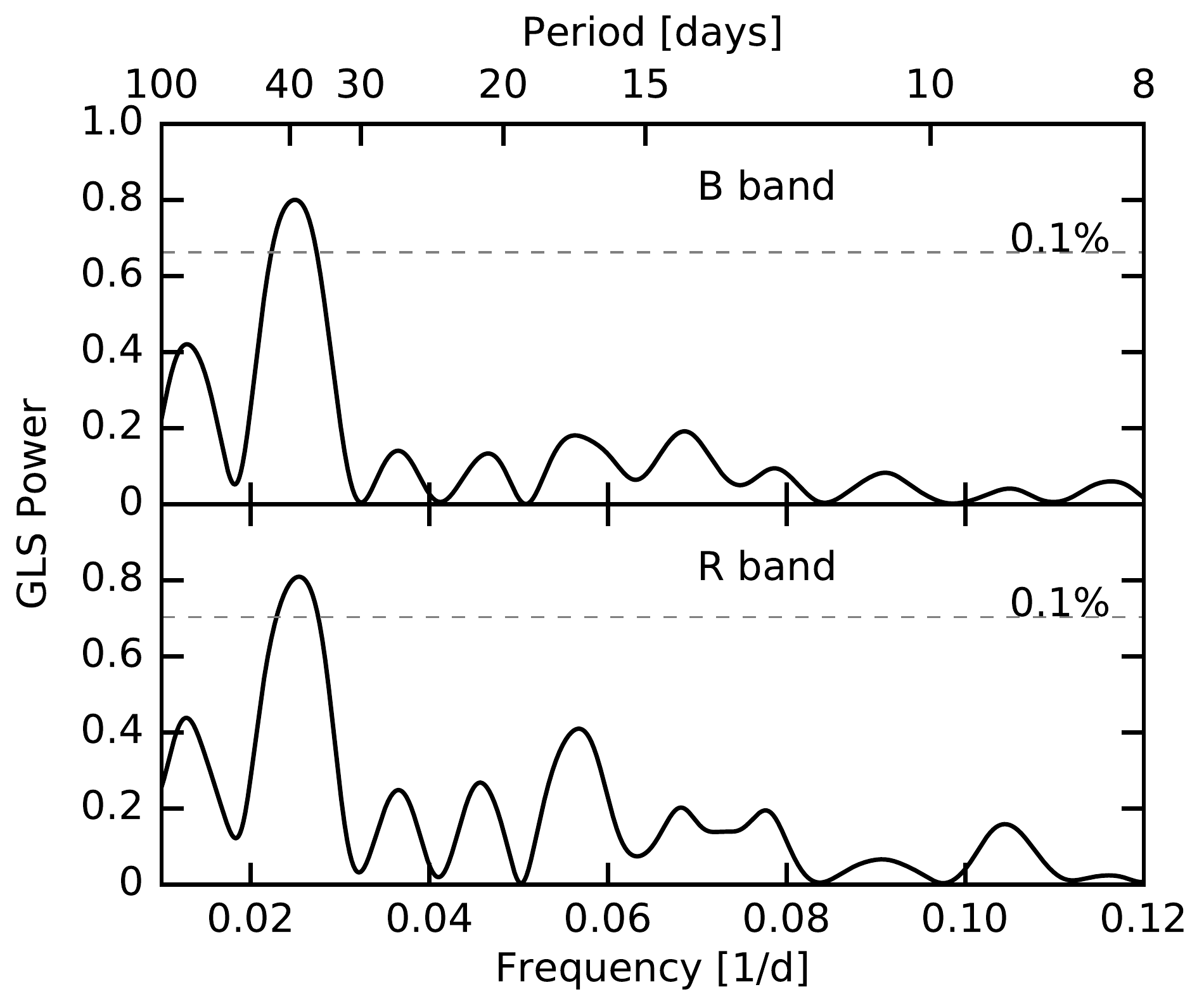}
	\caption{GLS periodogram of the {\it B} (upper panel) and {\it R} (lower panel) photometric data sets. The horizontal line indicates the 0.1\% FAP level. Both data sets show a significant peak at $\sim 40$ days indicating the stellar rotation period. 
	\label{fig:phot_periodogram}}
\end{figure}

To get a better estimate of the stellar rotation period,
we fit both bands simultaneously with a sine wave function
and forced both light curves to have the same frequency 
($f_{BR}$) and phase ($\phi_{BR}$),
but allowed the offsets ($\gamma_{B}$ and $\gamma_{R}$) 
and amplitudes ($A_{B}$ and $A_{R}$) to be different for each band. 
In total we fit for six parameters
($f_{BR}$, $\phi_{BR}$, $\gamma_{B}$, $\gamma_{R}$, $A_{B}$, and $A_{R}$)
and performed a Markov Chain Monte Carlo (MCMC)
using the \texttt{emcee} ensemble sampler 
\citep{dfm:2013}.
We adopted flat uniform priors for all parameters
and estimate the rotation frequency to be 
$0.02524 \pm 0.00032$ day$^{-1}$ ($39.63 \pm 0.50$ days).
This value is in agreement with the one estimated 
using the {\it K2} photometry
where \cite{Cloutier:2017} derived a value 
of $38.6^{+0.6}_{-0.4}$ days using Gaussian processes
and
\cite{Stelzer:2016} derived a value of 40.8 days using 
auto-correlation function (private comm.).

We estimated a photometric variability of 
$8.7 \pm 0.5$ mmag in {\it B} and 
a smaller variability of
$6.9 \pm 0.5$ mmag in {\it R}. 
This difference is expected when the photometric variability 
is due to cool spots,
since the contrast between the spots
and the photosphere decreases at redder wavelengths.
Figure~\ref{fig:phot_time_series} 
shows the photometric variations in the {\it B} filter (in blue)
and the {\it R} filter (in red) and the best fit model. 
In Tables~\ref{tab:data-photb} and \ref{tab:data-photr} 
we provide the differential photometry 
in {\it B} and {\it R} bands respectively.

\subsection{Spectroscopic Indicators} \label{sec:activity-spectro}

\begin{figure}[t!]
	\includegraphics[width=0.5\textwidth]{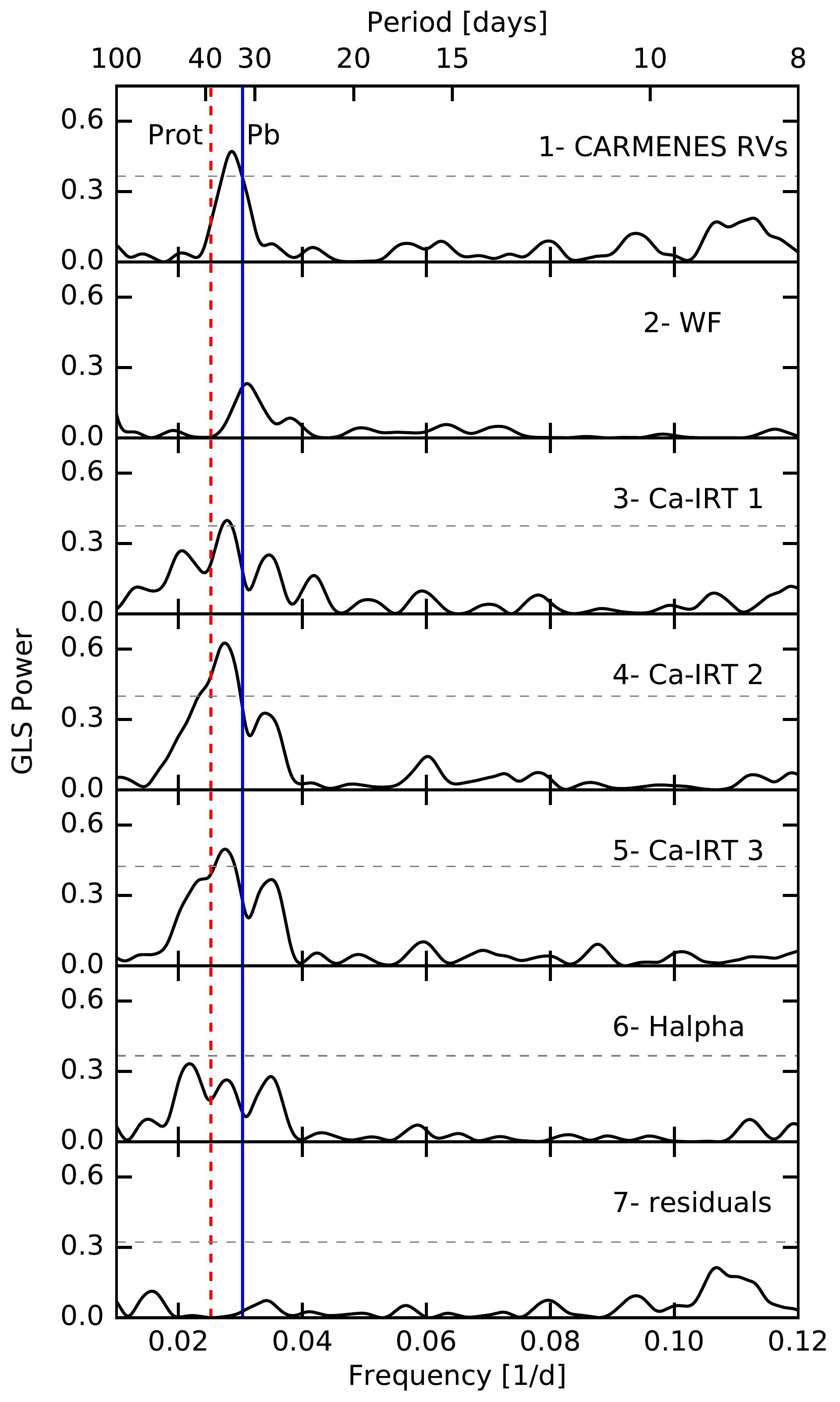}
	\caption{{\it From top to bottom}: GLS periodogram of the RVs, window function, the three Ca\,{\sc ii}~IRT lines, H$\alpha$ line, and the RV residuals. The blue solid line shows the orbital period of the planet, $P_b$, and the red dashed line indicates the stellar rotation period, $P_\mathrm{rot}$. The dashed horizontal lines show the 0.1\% FAP. Excess power in the RVs close to the orbital period of the planet indicates the presence of the RV signal of the planet in the data. Prominent peaks in the Ca\,{\sc ii}~IRT and H$\alpha$ lines hint at the rotation period of the star.
	\label{fig:rv_act_periodogram}}
\end{figure}

\begin{figure}[t!]
	\includegraphics[width=0.5\textwidth]{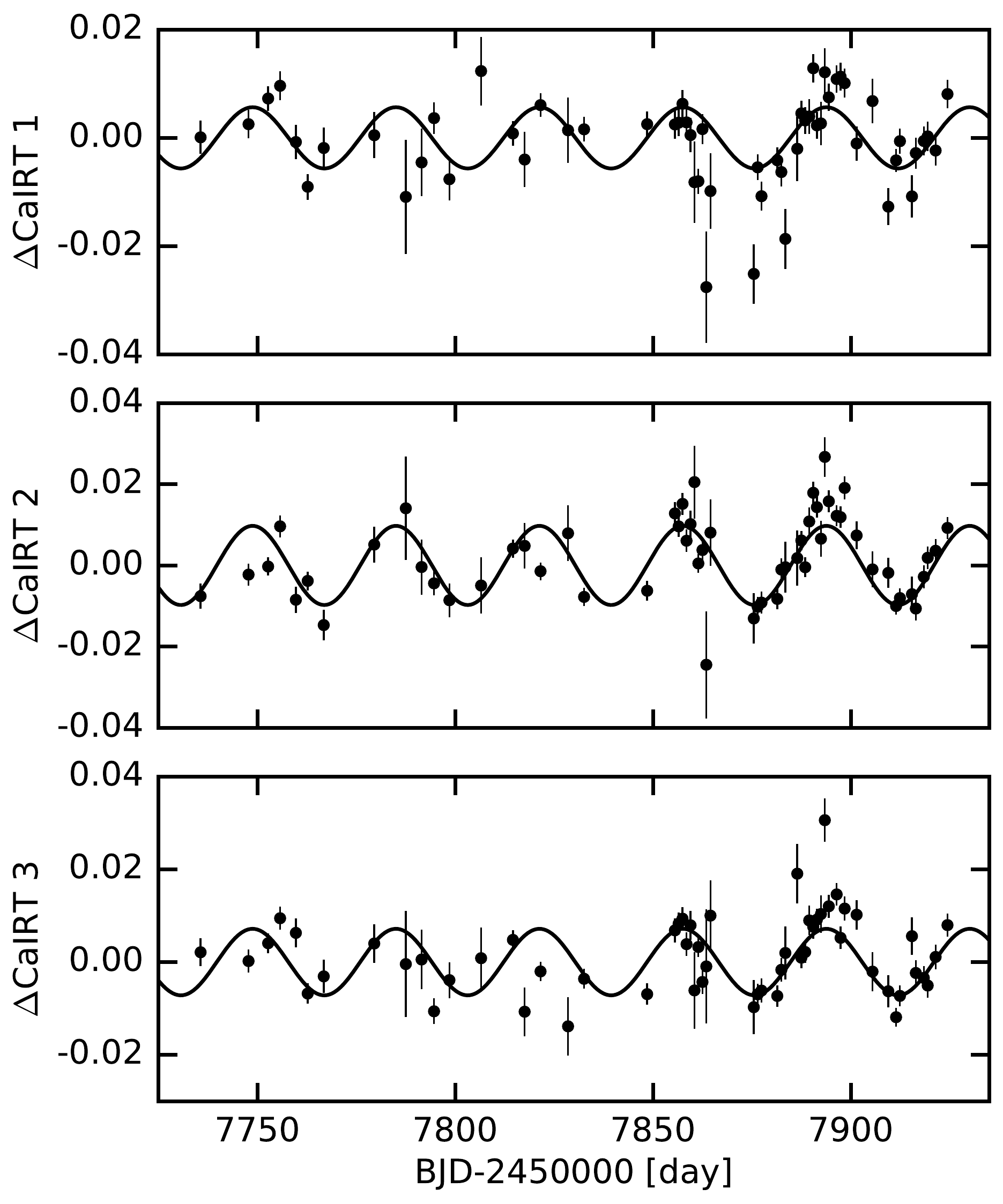}
	\caption{Time series of the three Ca\,{\sc ii}~IRT lines. The black curve shows the best fit to the data using a sinusoidal fit of which we estimate a period of $\sim$ 36 days.}
	\label{fig:cairt-time-series}
\end{figure}

The most common and widely used spectroscopic activity indicators 
can be divided into two different types:
the chromospheric and the photospheric ones.
The chromospheric activity indicators 
measure the excess of flux in the cores of e.g.
Ca\,{\sc ii}~H\&K, 
calcium infrared triplet
(Ca\,{\sc ii}~IRT), Na\,{\sc i}~doublet, and H$\alpha$ lines.
The core of these lines have their origin in the stellar chromosphere 
and hence they trace stellar magnetic activity.
The photospheric activity indicators 
measure the degree of asymmetry 
in the line profile.
The presence of spots on the photosphere 
distort the spectral lines
and therefore periodic variability
of the full width at half maximum (FWHM) and bisector span (BS)
of the 
cross-correlation function (CCF)
could indicate the presence of spots.
\cite{Zechmeister:2018} recently showed that 
the chromatic index is also an important photospheric indicator (see below).

The SERVAL pipeline provides the line indices of the
Ca\,{\sc ii}~IRT,
H$\alpha$, and Na\,{\sc i}~doublet. 
The three Ca\,{\sc ii}~IRT lines are 
centered at 8498.02 {\AA},
8542.09 {\AA}, and 8662.14 {\AA},
the H$\alpha$ line is centered at 6562.81 {\AA},
and the Na\,{\sc i}~D lines are centered at 5889.95 {\AA} and 5895.92 {\AA}.
The pipeline also computes the differential line width (dLW)
and the chromatic RV index.
The former is a measure of the 
relative change of the width of the average absorption line
and the latter is a measure of the wavelength dependency on the RV.
We refer the reader to \cite{Zechmeister:2018} for a detailed description 
of how the various activity diagnostics are computed.

We performed a period search analysis using GLS 
to search for a significant periodicity 
that could be related to stellar activity.
Figure~\ref{fig:rv_act_periodogram} (panels 3 - 6)
displays the periodograms
of the indicators that show a significant peak. 
Although we inspected a wide range of frequencies,
we only show the frequency range of interest
that covers the stellar rotation frequency, 
the planetary frequency of the transiting planet,
and the potential 9-day signal
(see Section~\ref{sec:planet9}).
All three Ca\,{\sc ii}~IRT indices
show a clear dominant peak at
$\sim 36$ days
with FAP $= 3\times 10^{-4}$, $< 10^{-4}$, and $= 10^{-4}$
for the Ca\,{\sc ii}~IRT~1, Ca\,{\sc ii}~IRT~2, and Ca\,{\sc ii}~IRT~3 lines
respectively, which was determined via bootstrap.
The H$\alpha$ periodogram shows three peaks
at 29, 36, and 45 days 
with FAP $=3.7\times 10^{-3}$ at 36 days.
The origin of the signal of both indicators
is consistent, 
within the frequency resolution,
with the rotational period of the star
derived from photometry (Section \ref{sec:activity-phot}).
Similar to the photometric data, 
we fit the Ca\,{\sc ii}~IRT indices simultaneously 
with a sine wave function forcing them to have the same 
frequency and phase, but allowed the offsets and amplitudes to vary.
Figure~\ref{fig:cairt-time-series} shows the 
Ca\,{\sc ii}~IRT line indices along with the 
best fit sinusoidal model. 
The Na\,{\sc i}~doublet and dLW periodograms, however, 
are free from significant peaks
even though the Na\,{\sc i}~lines were expected to be
good activity indicators 
for early M-dwarfs \citep{Silva:2011, Robertson:2015}.
We report the data of the activity indicators 
in Table \ref{tab:data-carmenes}.

In addition to the indicators provided by SERVAL,
we computed the CCF for each spectrum
by cross-correlating the spectrum
with a weighted binary mask
that was built 
by co-adding all the observed spectra of the star itself.
We selected around 4000 deep,
narrow, and unblended lines, 
which were weighted according to
their contrast and inverse FWHM.
We computed one CCF for each spectral order 
and the final CCF was computed 
by combining all the individual CCFs according to signal-to-noise.
A Gaussian function was fitted to the combined CCF.
From this, the FWHM and bisector span were derived.
A period analysis of the FWHM and bisector span 
does not show significant periods.
The lines in a typical M dwarf spectrum 
are blended and, thus, may mask changes in the FWHM and bisector span,
which could be the reason why 
these indicators do not show a variability.
Another reason is probably the low projected rotational velocity 
of the star ($v\sin i$).
\cite{Reiners:2017} imposed an upper limit on $v\sin i$ 
at $ 2 \, \mathrm{ km \, s^{-1}}$.
However, from the stellar radius and rotation period 
(Table \ref{tab:param}),
we estimate a true equatorial velocity
$v$ of only $ 0.53 \, \mathrm{ km \, s^{-1}}$.
The spot-BS relationship from \cite{Saar:1997}
predicts, for $v\sin i = 0.53 \, \mathrm{ km \, s^{-1}}$,
a bisector variability of 0.01 \mpersec, 
which is too small to measure.

%

\subsection{Photospheric vs Chromospheric Variations} \label{sec:chromoVSphoto}

\begin{figure}[t!]
	\includegraphics[width=0.5\textwidth]{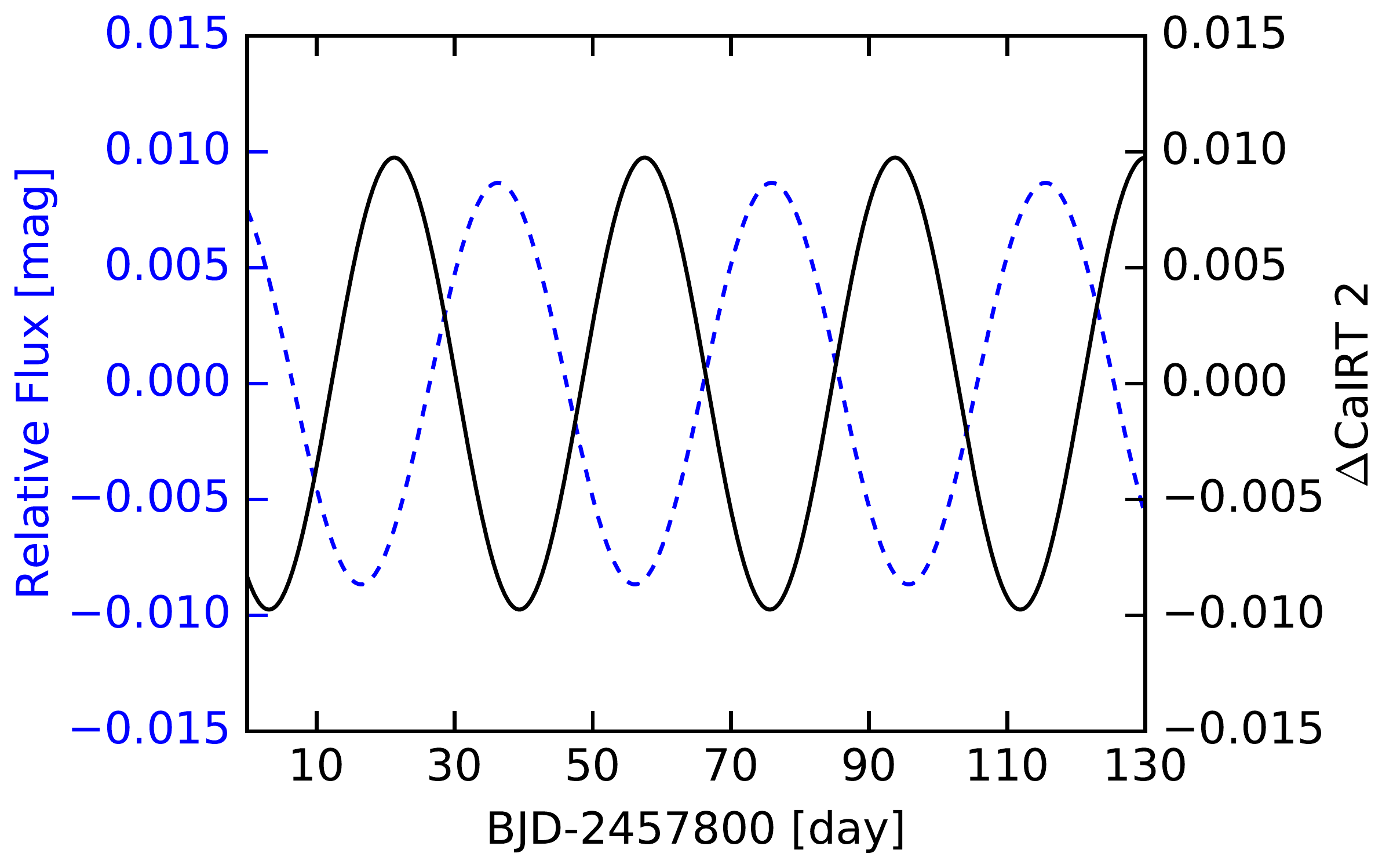}
	\caption{The blue dashed curve shows the photometric variability of K2-18 in the {\it B} band with a period of $\sim 40$ days. This is the same model shown in Figure~\ref{fig:phot_time_series}. The solid black curve is the best sine fit of the Ca\,{\sc ii}~IRT~2 line with a period of $\sim 36$ days. During this time interval, the two curves are 180$^\circ$ out-of-phase and show an anti-correlation between the photosphere and the chromosphere, especially in the second half of the data set.
	\label{fig:photo_vs_chromo}}
\end{figure}

The star shows photometric variability 
with a stellar rotation period of $39.63 \pm 0.50$ days.
The semi-amplitude is 0.87\% in the {\it B} band 
and 0.69\% in the {\it R} band.
K2-18 also shows chromospheric variability in 
the Ca\,{\sc ii}~IRT and H$\alpha$ lines with a period 
consistent with the rotation period derived from photometry
within the frequency resolution.
Figure~\ref{fig:photo_vs_chromo}
shows the variations of the 
Ca\,{\sc ii}~IRT second index and the best fit model
(solid black curve)
and the photometric variability
of K2-18 in the {\it B} band (dashed blue curve). 
There is an anti-correlation between the
photometric and the chromospheric variability.
The chromosphere shows variations
which are 180$^\circ$ out-of-phase with the photosphere.
Similar trends are seen with the first and third Ca\,{\sc ii}~IRT indices
and the H$\alpha$ line.
This demonstrates that for high Ca\,{\sc ii} emission values,
a minimum in the photometric light curve is observed. 
This is expected if active chromospheric regions 
are present on top of a photospheric spot. 
This is not the first time that an anti-correlation 
between the chromosphere and photosphere of M dwarfs is observed.
\cite{Bonfils:2007}
reported an anti-correlation 
for GJ~674, which is also an early M2.5 dwarf.
It would be worth checking whether the 
anti-correlation will hold for 
late M dwarfs.

We conclude that
K2-18 is a moderately active star
and
there is an anti-correlation 
between the photospheric and chromospheric variations,
which is consistent with the previous results of 
\cite{Radick:1998} for younger more active stars.
Finally,
although H$\alpha$ is a good activity indicator
\citep{Kurster:2003,Hatzes:2015,Robertson:2015,Jeffers:2018},
the Ca\,{\sc ii}~IRT lines show a significantly stronger peak 
compared to H$\alpha$. 
Ca\,{\sc ii}~IRT are thus good chromospheric 
activity proxies 
(see discussion by \cite{Martin:2017})
and 
provide a promising approach
to detect stellar activity signals in M dwarfs,
where the signal-to-noise is too low to 
measure Ca\,{\sc ii}~H\&K lines,
especially for mid and late M dwarfs.
This is also in agreement 
with the findings of \cite{Robertson:2016}.

\section{Radial Velocity Analysis} \label{sec:rv}

\subsection{Periodogram Analysis of the RVs}

\cite{Benneke:2017} analysed the {\it K2}
and {\it Spitzer} light curves
and derived an orbital period of 
$P = 32.939614^{+0.000101}_{-0.000084}$ days.
To ensure that we have detected the planet signal with high significance,
we performed a periodogram analysis for the RVs 
obtained with CARMENES-VIS. 
The RV measurements show a peak at 34.97 days with a FAP $< 10^{-4}$
(Figure~\ref{fig:rv_act_periodogram}, panel 1). 
This peak is approximately the mean of the 
planetary orbital frequency and the stellar rotation frequency
(0.02524 day$^{-1}$),
as measured in Section \ref{sec:activity-phot}.
The peak in the periodogram is therefore 
not centered at the orbital period of the planet 
but is shifted halfway 
between the stellar rotation frequency and the planetary orbital 
frequency. 
This shows that the RVs
are contaminated by stellar activity,
which is conceivable since the star 
is moderatively active (Section \ref{sec:activity}).

To assess the false alarm probability (FAP) of the planetary signal
and, hence, to confirm the detection of the planet,
we applied the bootstrap randomization technique.
Unlike the previous analysis where we
computed the GLS for the randomly shuffled data set
(see Section \ref{sec:activity-phot}),
this time we fitted an adapted model to the randomized data points.
The model 
employed the known ephemeris of the planet from \cite{Benneke:2017}, 
assumed zero eccentricity, 
and had only the RV semi-amplitude $K_b$ and the RV zero point (offset) 
as free parameters.
We performed this 100,000 times
and found that the FAP to infer a $K_b$ amplitude 
as large as (or larger than) the one 
estimated from the original data is $< 10^{-5}$
and the FAP to get a $\chi^2$ 
as small as (or smaller 
than) the one from the original fit
and finding at the same time that $K_b$ 
is positive is also $< 10^{-5}$.
This ensures that given the known ephemeris of the planet,
we are confident that there is a signal at the known ephemeris,
which can be a combination of the planet and activity signals.
In Section~\ref{sec:orbital-analysis}
we address several tests that we performed 
to check whether the RVs and, therefore, 
the planetary amplitude is affected by activity.

Signals that are sampled at discrete times 
can produce fake signals in the periodogram
that are due instead to observational patterns.
In order to check for periodicities due to sampling, 
we applied the GLS on the window function (WF), 
which is 
a periodogram analysis of the observation times. 
The GLS shows a peak at 32.2 days 
(Figure~\ref{fig:rv_act_periodogram}, panel 2)
which is very close to the orbital period of the planet. 
The reason for that peak is because we 
aimed to observe the star on a daily basis.
However, some nights were lost due to bad weather 
and more importantly 
during dark nights, roughly for a couple of lunar cycles,
another instrument was mounted on the telescope and 
no observations were carried out with CARMENES.
This pattern could have
caused the peak in the WF which is 
close to the lunar synodic cycle.

The presence of a peak in the WF 
would lead to the detection 
of the wrong frequency 
when there is a signal in the data.
\cite{Dawson:2010} showed 
that the reported periods of 55 Cnc\,e
and HD 156668\,b from their respective discovery papers
were actually wrong and affected by daily aliases.
In the case of \ktwo, 
first we have evidence that the star is moderately active 
(Section \ref{sec:activity})
and, as a result, we anticipate the presence 
of a signal in the RVs 
close to the stellar rotation frequency. 
Second, the planet transits 
\citep{Montet:2015, Benneke:2017}
and, therefore, we expect another signal in the data 
close to the orbital period of the planet. 
However, the proximity of the stellar rotation frequency 
to the planetary orbital one
makes separating them challenging,
since the frequencies are not resolved
given the time span of the data set.

Given the presence of the peak in the WF
and assuming the presence of one signal 
in the data 
(either the planetary signal
or the stellar rotational period),
is it possible to retrieve the signal 
at the right frequency?
To answer this question,
we generated a single synthetic sinusoidal signal 
sampled at times identical to the real RVs.
The uncertainty of every point
corresponded to the uncertainties derived from the RVs.
We generated two different sets, 
each with an amplitude of 3 \mpersec,
one set using the 
stellar rotation frequency and 
a second set using the planetary frequency.
Finally, for the synthetic data generated 
using the rotational frequency, instead of fixing the phase,
we covered a grid of phases
$\left[-\pi, -0.9\pi, ..., \pi \right]$.
For the planetary signal we assumed that 
the phase is well constrained.
We then did a periodogram analysis for each set
and could recover a peak at the true frequency.
This test shows that even though the WF shows a peak,
we can still retrieve the 
signal at the right frequency 
(planet frequency or the stellar rotation frequency)
given the data sampling.
Hence, the data set is not affected by aliases

\begin{figure*}[t!]
	\includegraphics[width=0.5\textwidth]{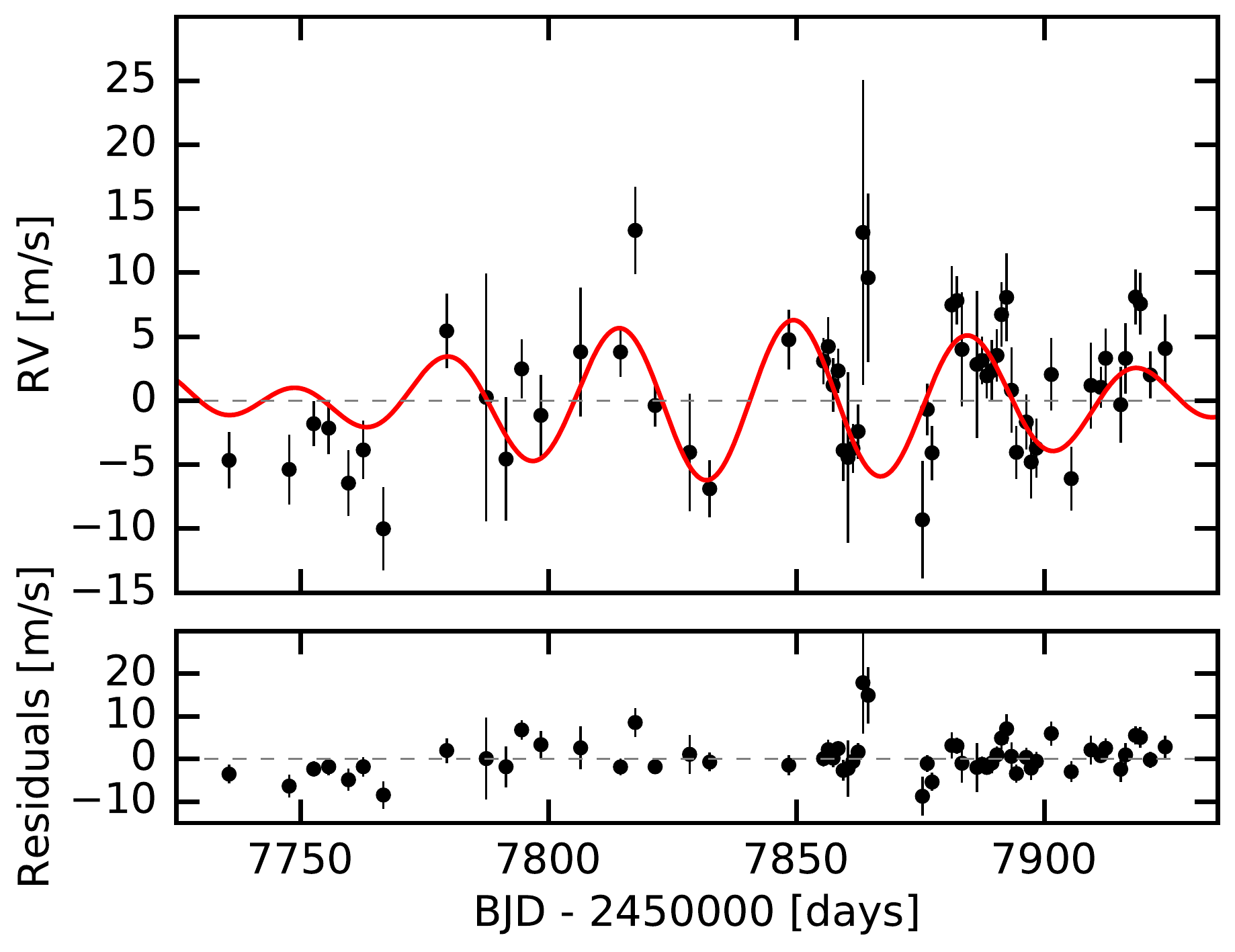}
	\includegraphics[width=0.5\textwidth]{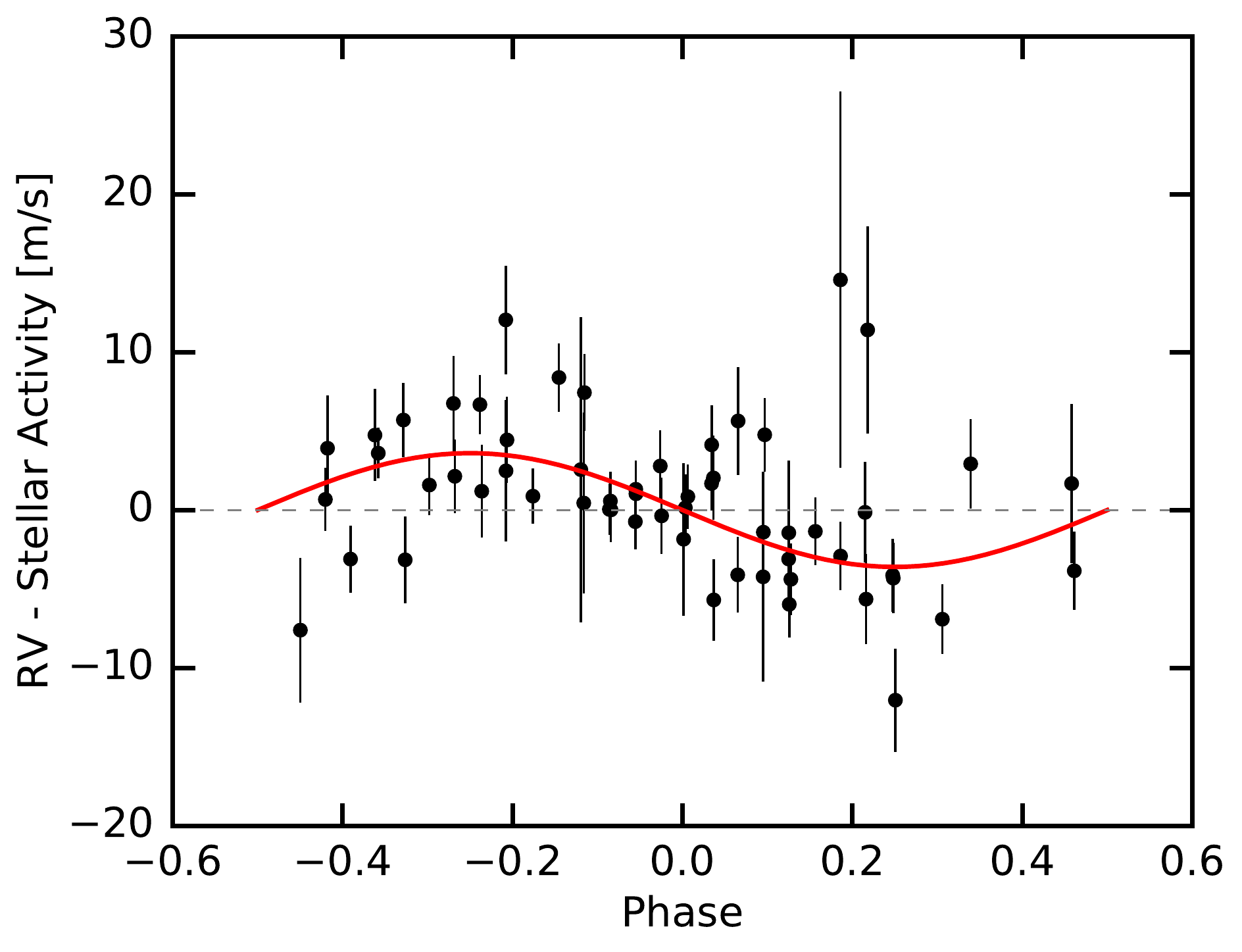}
	\caption{{\it Left:} CARMENES-VIS RVs modeled with a circular Keplerian signal of \ktwo\ plus stellar activity modeled with a periodic sine function (red line), and the residuals to the best-fit model. {\it Right:} Phase-folded activity-corrected RVs along with the best-fit planetary model. \label{fig:rvplot}}
\end{figure*}

In short, the planet's orbital period
is 32.94 days \citep{Benneke:2017}
and the stellar rotation period is $\sim 40$ days.
The RVs are not only affected by activity 
but the WF also shows a peak close to 32.2 days,
caused by observational patterns in the way the data was sampled. 
Previous studies 
\citep{Robertson-Mahadevan:2014,Vanderburg:2016} 
showed the difficulty in detecting RV planets
in orbits close to the stellar rotation period. 
\cite{Hatzes:2013} and \cite{Rajpaul:2016}
demonstrated that the WF 
can give rise to fake signals in the periodogram
that mimicked the presence of a planet 
around $\alpha$ Cen B which was reported by \cite{Dumusque:2012}.
In the case of \ktwo, the planet transits 
and hence its existence is undeniable. 
However, a closer look at the WF is needed 
to check whether the RV signal of the planet 
is present in the data.
This case demonstrates the difficulty in detecting 
non-transiting low-mass planets
not only at orbits close to the stellar rotation period
but also when observational patterns are present in the data.


\subsection{Orbital Analysis of \ktwo} \label{sec:orbital-analysis}

We performed joint modeling
of the photometric light curves obtained with STELLA 
and the RV measurements. 
Similar to Section \ref{sec:activity-phot},
we modeled the photometric data of both bands 
with a sine wave function 
and fit for 
$f_{BR}$, $\phi_{BR}$, $\gamma_{B}$, $\gamma_{R}$, $A_{B}$, and $A_{R}$.
We adopted uniform priors for the phase and offsets 
of the stellar photometric variability.
For $f_{BR}$, $A_{B}$, and $A_{R}$ we adopted Gaussian priors
centered at $0.02524$ day$^{-1}$, 8.7 mmag, and 6.9 mmag, respectively
and with a standard deviation of 0.00032 day$^{-1}$ and 0.5 mmag for both amplitudes
(see Section~\ref{sec:activity-phot}).
We fit the RV measurements with a Keplerian model 
assuming a circular orbit ($e$ = 0)
and using the combined {\it K2} and {\it Spitzer} 
ephemeris, i.e. we fixed
the mid-transit time $T_0$ and $P_b$
to the values derived photometrically by \cite{Benneke:2017}
since these parameters are tightly constrained.
We accounted for stellar activity in the RV data by 
assuming that it has a sinusoidal function 
whose frequency is constrained 
from the photometric light curves. 
We let the phase of the stellar activity $\phi_\mathrm{act}$ free,
and thus fit for the phase, 
amplitude $K_\mathrm{act}$, and frequency $f_{BR}$
of the stellar activity.
We adopted non-informative priors 
for the offset, $\phi_\mathrm{act}$, $K_\mathrm{act}$, and $K_b$.
The joint analysis was then performed
using \texttt{emcee} \citep{dfm:2013}
and in total we fit for 10 parameters:
6 parameters for the photometric data
(mentioned above)
and 5 parameters for the RV data
($\gamma$, $K_b$, $\phi_\mathrm{act}$, $K_\mathrm{act}$, and $f_{BR}$);
the stellar rotation frequency 
is the same in both data sets.

The best fit model gave a planetary semi-amplitude 
of $K_b = 3.60^{+0.53}_{-0.51}$ \mpersec\
and a stellar activity semi-amplitude of
$K_\mathrm{act} = 2.72 \pm 0.50$ \mpersec,
corresponding to a planetary mass of 
$M_b = 9.07^{+1.58}_{-1.49} \, M_{\oplus} $,
using $M_* = 0.359 \pm 0.047 $ $M_{\odot}$.
Figure~\ref{fig:carmenes-corner-plot} 
shows the joint and marginalized posterior
constraints on the model parameters.
Using the transit depth , $R_\mathrm{b}/R_*$, 
and stellar radius, $R_*$, 
as reported in \cite{Benneke:2017} and provided in 
Table~\ref{tab:param}, 
we derive a planetary radius
$R_b = 2.37 \pm 0.22 \, R_{\oplus} $\footnote{Given the 10\% measurement
uncertainty on the stellar radius, 
we expect a 10\% measurement uncertainty on the planetary radius. 
However, \cite{Benneke:2017} reported
a value on the order of 1\%.}, 
this corresponds to a planetary density of 
$\rho_b = 4.18^{+1.71}_{-1.17}$ g cm$^{-3}$.
The $v\sin i$ and spot filling factor 
estimated from photometry
yield an RV semi-amplitude of 2.7 \mpersec\
for spots using the relationship by \cite{Hatzes:2002},
which is in excellent agreement with the one estimated 
using the RV data.
The planetary semi-amplitude value
is consistent with the one derived 
using HARPS RVs
by \cite{Cloutier:2017} at the 1-sigma level.
The best fit model and the phased RVs 
are shown in Figure~\ref{fig:rvplot}.
We report 
the stellar and planetary parameters
used in this study
and the median values of all the parameters,
along with the $16^{\mathrm{th}}$ and $84^{\mathrm{th}}$
percentiles of the marginalized posterior distributions
in Table \ref{tab:param}.

\begin{figure}[t!]
	\includegraphics[width=0.5\textwidth]{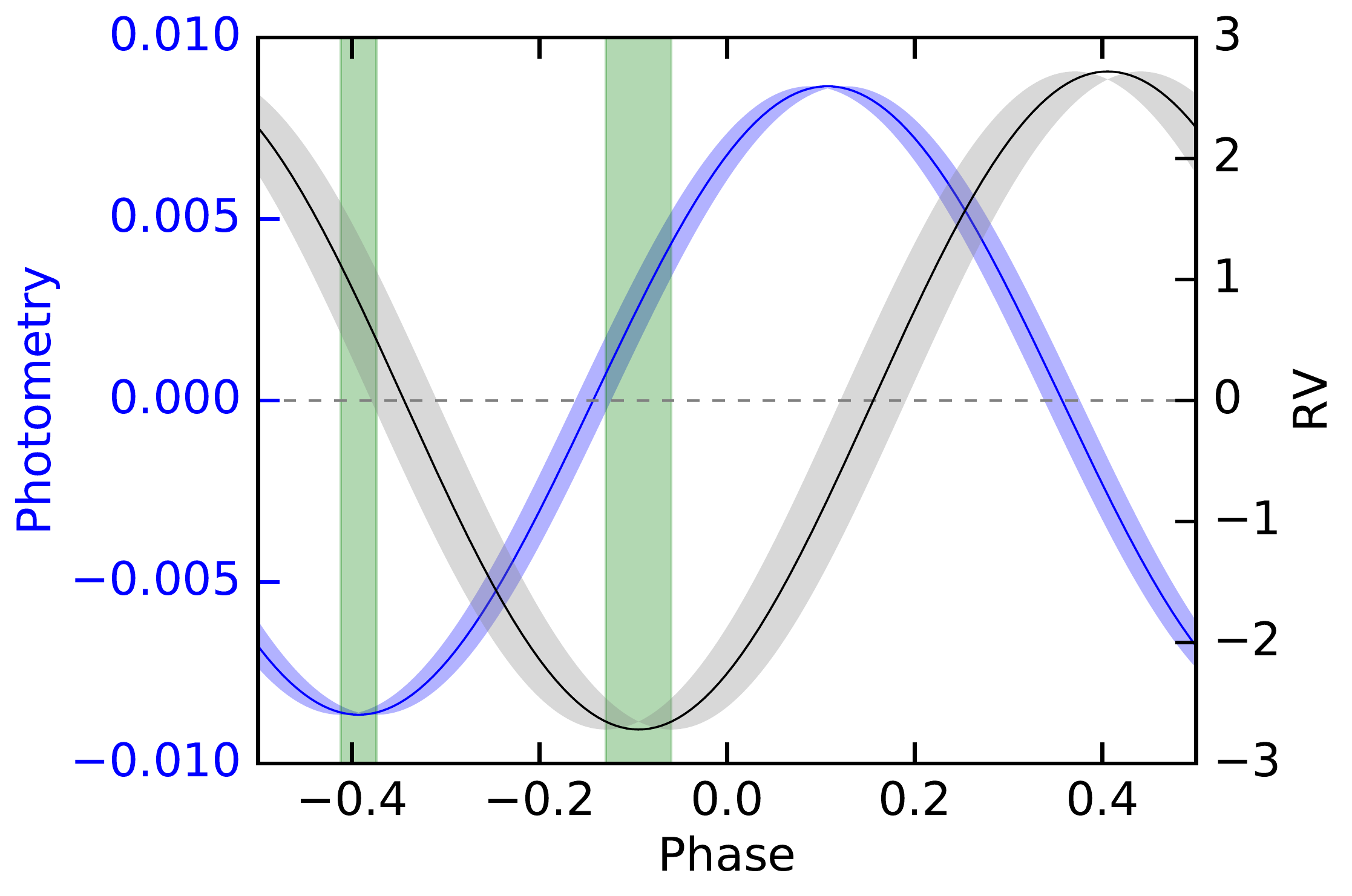}
	\caption{Phase-folded photometric light curve in the {\it B} band (blue) and RV signal due to activity (black), along with their 1$\sigma$ uncertainties. To aid the eye, the minima of both curves are shaded in green. Within the error bars the phase shift between the two curves is 90$^\circ$, as expected if the photometric and RV signals are due to cool spots crossing the visible stellar surface as the star rotates.
	\label{fig:phaseshift}}
\end{figure}

To further test whether the activity signal 
is due to cool spots,
we compared the phase shift between the photometric light curve 
and the RV signal due to activity.
Figure~\ref{fig:phaseshift}
shows the phase folded photometric light curve 
in the {\it B} band in blue and
the RV signal in black.
When the spot is at the center of the stellar surface
(minimum in the photometric light curve),
the contribution of the spot to the RV signal is close to zero.
As the spot moves along the stellar surface
to the receding redshifted limb 
(zero in the photometric light curve), 
the star appears to be blueshifted (minimum in the RV curve).
Therefore, the phase shift is $\sim 90^\circ$.
This is expected if the variations are due to cool spots,
which is also consistent with the multi-wavelength photometry
analysis (Section \ref{sec:activity-phot}).
This is only considering the \textit{flux} effect of dark spots.
In general, the RV variations in active regions are induced 
by two different physical processes:
first 
the asymmetry in the stellar line profiles created by star spots
and second the suppression of the convective blueshift 
effect due to the presence of strong magnetic fields
that inhibit convection inside active regions.
The convective blueshift effect could explain why 
the RV curve appears shifted a bit vertically at the 
minimum phase of the photometric lightcurve.

Even though the star shows periodic 
photometric variability,  
there is evidence that the chromosphere
does not show strict periodic sine-like variability 
(see Section \ref{sec:planet9} and Figure~\ref{fig:cairt-time-series},
where some points deviate from the best fit curve,
especially Ca\,{\sc ii}~IRT 1 and Ca\,{\sc ii}~IRT 3).
Therefore, modelling the RV signal of stellar activity 
by a periodic sinusoidal function
might not be the best approach.   
However, 
we next argue that the derived planetary semi-amplitude 
is not dependent on our choice 
of the model used to account for stellar activity.
We performed several tests to check this dependency. 
First, following \cite{Baluev:2009}, 
we accounted for stellar activity by adding 
a constant white noise term often referred to as
the RV jitter term, $\sigma_\mathrm{jitter}$.
The jitter term is treated as an additional source of 
Gaussian noise with variance $\sigma^2_\mathrm{jitter}$
and is added
in quadrature to the estimated RV uncertainties
\citep{Ford:2006}.
We derived a planetary semi-amplitude $K_b = 3.38^{+0.75}_{-0.76}$ \mpersec\
and an RV jitter
$\sigma_\mathrm{jitter} = 3.02^{+0.57}_{-0.53}$ \mpersec.
The planetary semi-amplitude derived using this model
is in agreement with the one derived 
previously,
within the 1-sigma error bars.

Second, to check whether the RVs 
are affected by stellar activity,
we looked for correlations 
between the raw RVs
and the various activity indicators mentioned in Section \ref{sec:activity-spectro}.
The upper panels in 
Figure~\ref{fig:rv-act-corr} in the Appendix~\ref{sec:appendix}
show the measured RVs
plotted against the activity indicators
and color coded according to the stellar rotational phase.
We did not find a linear correlation between 
any of these quantities and the measured RVs.
However, there is a slight indication that the color coded 
data points follow a circular path,
especially for Ca\,{\sc ii}~IRT 2,
but not with high significance.
We further repeated the same analysis 
after the removal of the planetary signal
and still did not find any significant correlations 
with the activity indicators. 
The results are shown in the lower panel of Figure~\ref{fig:rv-act-corr}.
Despite detecting a signal close to the 
stellar rotational period in both the RVs and the Ca\,{\sc ii}~IRT lines,
no evident linear or circular correlation is seen,
indicating that the relation is quite complex.

Third, we ignored activity and fit the RVs with a single Keplerian signal
and fixed $T_0$ and $P_b$ to the known photometric values.
We estimated a planetary semi-amplitude 
$K_b = 3.35 \pm 0.47$ \mpersec\ which is also 
in agreement with the previous results.
We further divided the data set into two,
each containing 29 data points,
and repeated the same analysis for the first and second half 
of the data.
We found similar planetary semi-amplitudes in both cases
and the values are given in Table~\ref{tab:test-act}.

As a final test\footnote{ 
\cite{Cloutier:2017} demonstrated that 
the planetary semi-amplitude derived by implementing a Gaussian Process model
(Model 1 in their Table~2) is consistent 
at the 1-sigma level with the model 
that neglects any contribution from stellar activity 
(their Model 4).
Also the covariance amplitude is in agreement with zero within the error bars 
$0.1_{-0.1}^{+2.8}$ \mpersec.},
we looked at the RV measurements
in the red and blue orders of CARMENES-VIS.
If the RVs are dominated by activity
due to active regions on the stellar surface,
then the planetary semi-amplitude in the blue 
part of the spectrum should be more affected by activity
whereas the red part should be less affected. 
As a result, if the star is active, 
a single Keplerian fit to the data 
should yield different planetary semi-amplitudes
for different orders. 
The RV measurements for K2-18 are available at 42 orders. 
We calculated an RV weighted mean average
for the first and second half of the orders,
which we will refer to as the {\it blue} RVs 
and as the {\it red} RVs respectively
and are reported in Table~\ref{tab:data-carmenes}.
The blue orders cover the wavelength range 
from 561 to 689 nm,
whereas the red orders cover the range 
from 697 to 905 nm.
We also ignored activity and fit separately 
the blue and red RVs with a 
Keplerian model with $T_0$ and $P_b$ fixed. 
We did this analysis for the full CARMENES-VIS data set, 
the first half, and the second half. 
So in total we repeated this analysis six times,
all of which yielded similar planetary semi-amplitudes 
within the error bars.
The values are reported in Table \ref{tab:test-act},
where we denote the original full wavelength coverage
RVs as full-$\lambda$ RVs.
We conclude that the RVs are not dominated by stellar activity, 
and that the estimation of the planetary semi-amplitude 
is robust 
and does not depend on the choice of model
used to account for stellar activity.

We also computed the results of Table~\ref{tab:test-act}
using a Keplerian model plus a sinusoidal model
to account for activity,
where we fit for the stellar rotation frequency.
We find that the planetary semi-amplitude 
is consistent within 1-sigma 
when computed for the 
full data, first half, and second half 
for the full spectral coverage, the red, and blue orders 
with one exception, 
the planetary amplitude computed 
for the second half in the red order. 
However, the value is in agreement at the 2-sigma level. 
Even though we expect the activity semi-amplitudes 
to be different in different orders, 
the semi-amplitudes derived 
are consistent either at the 1-sigma or at the 2-sigma level. 
This could be explained by 
the low amplitude signals in both order ranges,
which are on the order of $2.7 \pm 0.73$ \mpersec\
i.e. a higher precision would be required 
to differentiate between the activity semi-amplitudes in different orders.

\begin{table}[t!]
\caption{%
The planetary semi-amplitudes $K_b$ derived for the full, first, and second half of the data set using 
the full-$\lambda$ RVs, the blue RVs, and the red RVs. 
\label{tab:test-act}}
\begin{center}
\begin{tabular}{cccc}
\toprule
$K_b \, (\mpersec)$    &  Full Set         & First Set         & Second Set \\
\hline
Full-$\lambda$ RVs &  3.35 $\pm$ 0.47  & 3.23 $\pm$ 0.66   & 3.10 $\pm$ 0.68      \\
Blue RVs               &  3.46 $\pm$ 0.55  & 3.71 $\pm$ 0.79   & $2.71^{+0.80}_{-0.77}$ \\
Red RVs                &  3.29 $\pm$ 0.46  & 2.77 $\pm$ 0.65   & 3.44 $\pm$ 0.64       \\
\hline
\end{tabular}
\end{center}
\end{table}

\subsection{Search for a Second Planet} \label{sec:planet9}

\begin{figure*}[t!]
	\includegraphics[width=\textwidth]{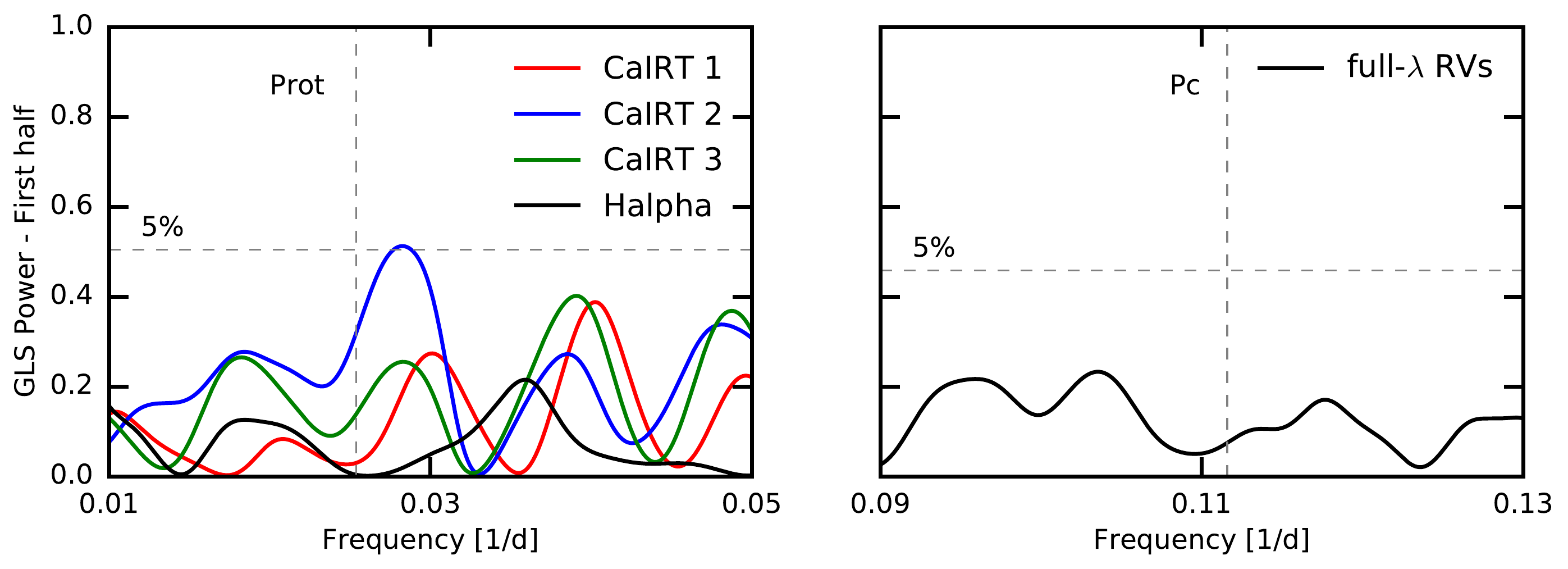}
	\includegraphics[width=\textwidth]{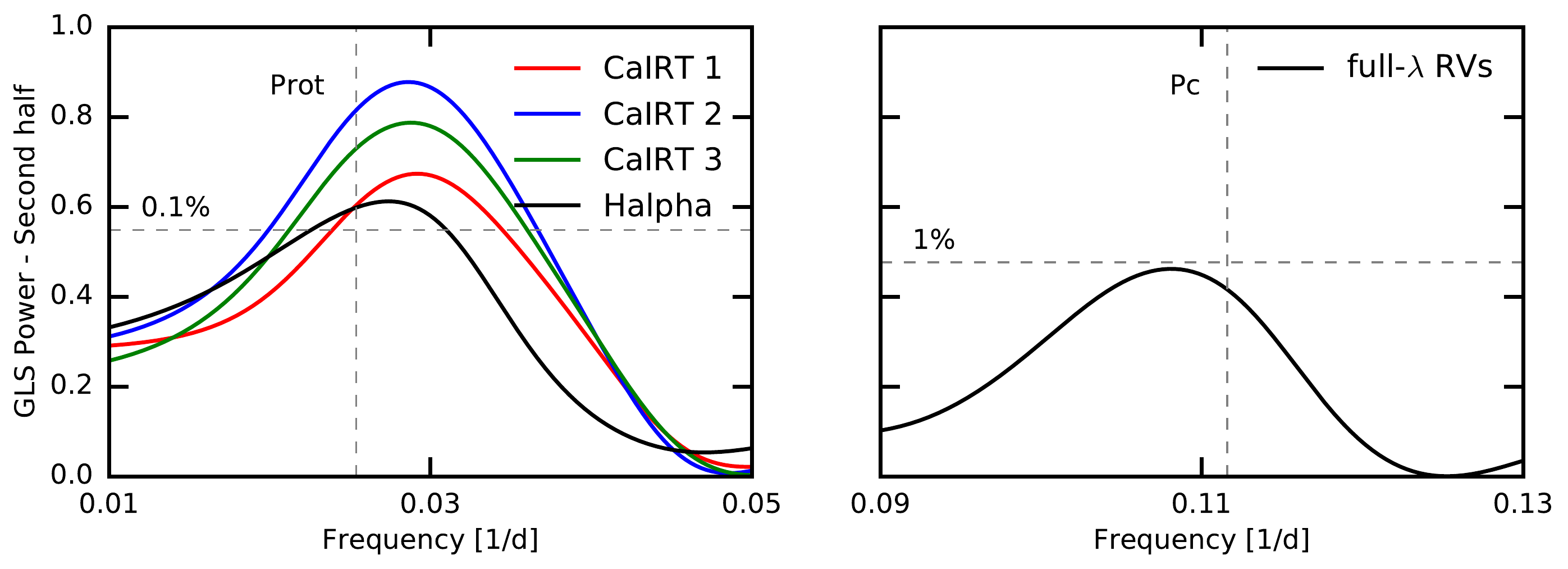}
	\caption{Periodograms of the first (top panels) and second (bottom panels) halves of the Ca\,{\sc ii}~IRT and H$\alpha$ lines (left panels) and CARMENES-VIS RVs (right panels). The dashed lines on the left and right show the stellar rotational period $P_\mathrm{rot}$ and the claimed period of the inner planet $P_c$ respectively. The signal of the inner planet is only present in the second half of the data set when all the spectroscopic indicators show a single significant signal at $P_\mathrm{rot}$.
	 \label{fig:innerplanet-act-corr}}
\end{figure*}

\begin{figure*}[t!]
	\includegraphics[width=\textwidth]{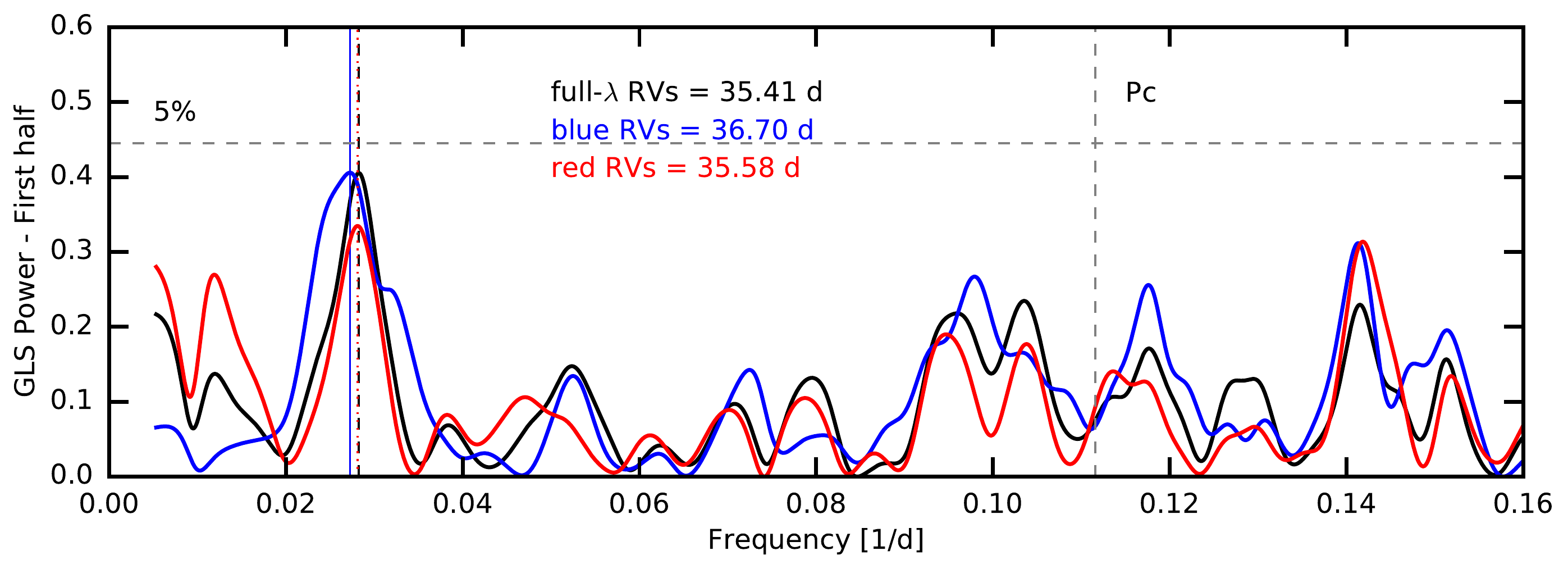}
	\includegraphics[width=\textwidth]{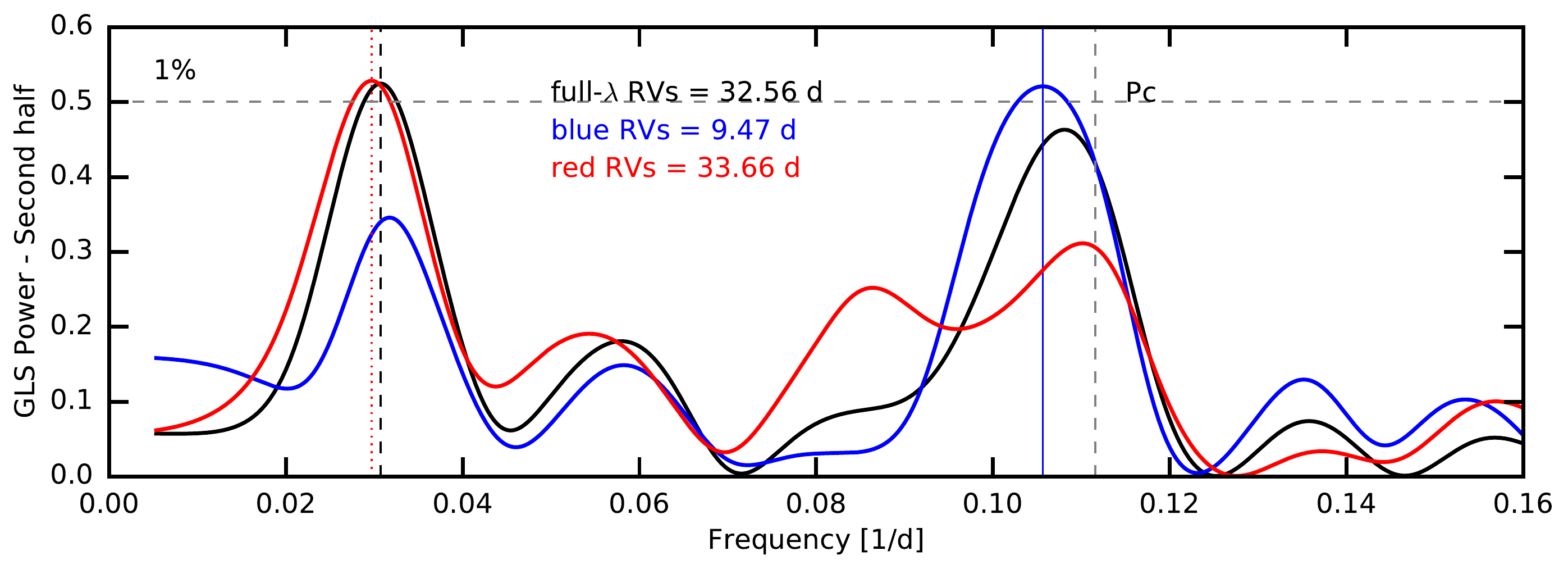}
	\caption{Periodograms of the first (top panel) and second (bottom panel) halves of the data set of the full-$\lambda$ RVs, blue RVs, and red RVs. The dashed, solid, and dotted lines indicate the peak with the highest GLS power for the full-$\lambda$ RVs, blue, and red RVs, respectively. The signal of the inner planet $P_c$ is only prominent in the second half of the data set when the star shows high level of activity.
	\label{fig:innerplanet-color}}
\end{figure*}

\cite{Cloutier:2017} used 75 HARPS RV measurements 
spanning approximately three seasons of observations to
estimate the mass of \ktwo\
and to search for additional planetary signals.
They reported a non-transiting planet, \ktwoc,
with a period of $8.962 \pm 0.008$ days
and a semi-amplitude of $4.63 \pm 0.72$ \mpersec.
The signal of \ktwoc\ is stronger than \ktwo\
(see Figure 2 in \cite{Cloutier:2017}).

We searched for the signal of the second planet 
in the CARMENES-VIS data set.
As mentioned in Section \ref{sec:rv}, 
the periodogram only shows
one strong peak at 34.97 days, 
the combined signal of the $\sim 33$-day period planet and 
the stellar rotation period.
The second strongest peak is around 9 days
with a FAP $>$ 5\% 
and significantly weaker than in the HARPS data.
We then subtracted the signal of the 33-day period planet 
and stellar activity
from the RVs and performed again a period analysis.
We still did not find a strong signal 
at the period of the supposed second (inner) planet
(Figure~\ref{fig:rv_act_periodogram}, panel 7).

In order to examine whether the absence of the 9-day signal
in the CARMENES-VIS data set is due to bad sampling,
we generated a synthetic RV data set 
assuming that there are two planets in the system
and using the real observing times of CARMENES.
We set the values of the 
orbital period, semi-amplitude, and time of inferior conjunction 
of both planets as derived by \cite{Cloutier:2017}:
$P_b = 32.93963$ days, $P_c = 8.962$ days,
$K_b = 3.18$ \mpersec,  $K_c = 4.63$ \mpersec,
$T_{0,b} = 2457264.39157$ BJD, and $T_{0,c} = 2457264.55$ BJD.
We further assumed that the uncertainty is the sum 
of the observational error and a random noise 
(drawn from a normal distribution centered at 0 
and a standard deviation of 0.25 \mpersec)
to attribute to the stellar jitter determined by \cite{Cloutier:2017}.
We then did a periodogram analysis and could recover an extremely strong
peak at 8.98 days with a FAP $< 0.1\%$.
This shows that our analysis is not affected by poor time sampling.

We also examined whether the 9-day signal 
could be caused by stellar activity,
since the period is 
near the fourth harmonic of the stellar rotation period
\citep[$39.63$ days -- Section \ref{sec:activity};][]{Cloutier:2017}.
We divided the full CARMENES-VIS data set into two, 
each consisting of 29 data points,
and did a periodogram analysis for each set
of the RVs, Ca\,{\sc ii}~IRT, and H$\alpha$ lines.
Figure~\ref{fig:innerplanet-act-corr}
shows the periodograms for both data sets.
The upper left and upper right panels 
show the periodograms of the activity indicators and 
RVs respectively for the first half of the CARMENES-VIS data set.
Similarly, the lower panels
show the periodograms for the second half of the data set. 
The dashed line in the periodograms of the activity indicators 
shows the stellar rotation period, $P_\mathrm{rot}$,
while
the dashed line in the RV periodograms indicates
the period of the inner planet, $P_c$, as estimated by \cite{Cloutier:2017}.
Note that for the activity indicators
only the periodogram region near the rotation period is shown, 
whereas for the RVs only the region around the 9-day signal is displayed.
The different levels of FAPs are indicated in the plot.
The first half of the RV data set 
does not show a power at the orbital period of the supposed inner planet.
That is also true when the Ca\,{\sc ii}~IRT and H$\alpha$ lines
do not show a consistent peak.
The second Ca\,{\sc ii}~IRT index
is the only indicator that shows a somewhat stronger peak
with a FAP $\sim 1$\%.
The other indicators do not show a prominent peak 
and notably H$\alpha$ shows no power at the
stellar rotation period.
The signal of the 9-day period 
appears only in the second half of the RV data set,
which occurs at the same time
when all the 
Ca\,{\sc ii}~IRT and H$\alpha$ lines 
show a prominent peak at the stellar rotation period
with a FAP $<$ 0.1\%,
demonstrating that the level of activity increased 
in this set.
This indicates that the signal of the 
9-day planet is absent when the star
is less active
and is present only when the 
level of activity increases significantly.
We thus conclude that the presence of the 9-day signal
correlates with the Ca\,{\sc ii}~IRT and H$\alpha$ lines.

This is further illustrated 
in Figure~\ref{fig:innerplanet-color},
which shows the periodograms of the
full-$\lambda$ RVs
and the 
blue and red RVs of CARMENES-VIS, 
which are calculated
as explained 
in Section \ref{sec:orbital-analysis}.
The periodogram 
for the blue RVs is shown in blue, 
for the red RVs in red, 
and for the full-$\lambda$ RVs in black. 
The legend indicates the period with the highest power 
for the different sets of RVs. 
The blue, red, and full-$\lambda$ RVs
show a single peak in the 
first half of the data set (upper panel)
close to 36 days. 
In the second half, 
interestingly the periodogram of the blue RVs
shows the highest GLS power close to 9 days,
while the red and full-$\lambda$ RVs
show the highest power close to the orbital period 
of the 33-day period planet.
This further demonstrates
that when the level of stellar activity increased, 
the blue RVs show a period 
at the fourth harmonic of the stellar rotation period
while the red RVs do not.
This is {in line with the notion that} RV variations 
due to photometric star spots are wavelength dependent 
and more prominent in the blue part of the spectrum,
while the variations get smaller at redder wavelengths
\citep{Reiners:2010}.
On the other hand, the RV variation
of a planetary signal is wavelength independent 
and should be constant at all wavelengths. 
This shows the importance of multi-wavelength RV measurements
to differentiate planetary from stellar activity signals.

\begin{table*}[t!]
\caption{Stellar and planetary parameters for the system K2-18.
\label{tab:param}}
\begin{center}
\begin{tabular}{lccc}
\toprule
Parameter &  & Value &  \\
\hline
Stellar parameters              &                   &     & \\
$ \qquad P_\mathrm{rot} $ [days] & & 39.63 $\pm 0.50$   &  \\
$ \qquad M_*$ [$M_{\odot}$] \tablenotemark{a}  &  & 0.359 $\pm 0.047$ &   \\
$ \qquad R_*$ [$R_{\odot}$] \tablenotemark{a}  &  & 0.411 $\pm 0.038$ &   \\
$ \qquad T_*$ [K]           \tablenotemark{a}  &  & 3457  $\pm 39$    &   \\
$ \qquad$[Fe/H] [dex]       \tablenotemark{a}  &  & 0.12 $\pm 0.16$ &    \\
\hline
Transit parameters               &                   &      &  \\
$ \qquad R_\mathrm{b}/R_* $ [\%] \tablenotemark{a} &  & 5.295$^{+0.061}_{-0.059}$ & \\
$ \qquad T_0 $ [BJD]  \tablenotemark{a}   & & 2457264.39144$^{+0.00059}_{-0.00066}$ &   \\
$ \qquad P_b $ [d]      \tablenotemark{a}   & & 32.939614$^{+0.000101}_{-0.000084}$ &   \\
$ \qquad R_\mathrm{b}$ [$R_{\oplus}$] \tablenotemark{b} & & 2.37$ \pm 0.22$  &  \\
$ \qquad i $ [deg]   \tablenotemark{a}  & & 89.5785$^{+0.0079}_{-0.0088}$       &   \\
\hline
\hline
 &         &  Models &  \\       
\hline
 & Planet only & Planet + sine & Planet + jitter   \\
Radial Velocity parameters                   &                   &     &   \\
$ \qquad K_b$ $\left[\mpersec\right]$ & 3.35 $\pm 0.47$ & $3.60^{+0.53}_{-0.51}$ & $3.38^{+0.75}_{-0.76}$  \\
$ \qquad K_\mathrm{act}$ $\left[\mpersec\right]$ & ... & $2.72 \pm 0.50$ & ...  \\
$ \qquad \sigma_\mathrm{jitter}$ $\left[\mpersec\right]$ & ... & ... & $3.02^{+0.57}_{-0.53}$ \\
$ \qquad e$                     & 0 (fixed)         & 0 (fixed)  & 0 (fixed)  \\
\hline
Planet parameters               &                   &      &  \\
$ \qquad a $ [au]     \tablenotemark{a} & 0.1429$^{+0.0060}_{-0.0065}$ & 0.1429$^{+0.006}_{-0.0065}$ & 0.1429$^{+0.006}_{-0.0065}$ \\
$ \qquad M_\mathrm{b}$ [$M_{\oplus}$]  & 8.43$^{+1.44}_{-1.35}$     & 9.06$^{+1.58}_{-1.49}$ & 8.49$^{+2.08}_{-1.97}$  \\
$ \qquad T_{\mathrm{eq}}$ [K]          & $283 \pm 15$ & $283 \pm 15$ & $283 \pm 15$  \\
$ \qquad \rho_{\mathrm{b}}$ [g cm$^{-3}$] & $3.89^{+1.58}_{-1.08}$  & $4.18^{+1.71}_{-1.17}$ & $3.90^{+1.77}_{-1.24}$  \\
\hline
\end{tabular}
\tablenotetext{a}{Parameters based on \cite{Benneke:2017}.}
\tablenotetext{b}{Recalculated the value using $R_\mathrm{b}/R_*$ and $R_*$ as derived by \cite{Benneke:2017}.}
\end{center}
\end{table*}

Notably, 
in the second half of the data set, 
when the star is relatively more active, 
the red and full-$\lambda$ RVs
show peaks 
much closer to the
the orbital period of the planet and are 
not shifted in value toward that of the stellar rotation.
It seems the contribution of activity 
to the RVs appears near the fourth harmonic
of the stellar rotation period 
and this set shows a clean planetary signal.

We conclude that, although we found evidence 
of the second planet signal announced recently 
by \cite{Cloutier:2017},
the peak is not significant in the 
CARMENES-VIS data set with a FAP $>$ 5\%.
The signal is also time and color variable
and correlates with stellar activity.
Given the sampling and the time baseline
of our observations, 
we conclude we do not have enough 
evidence to confirm the presence of the second 
inner planet
and there is a strong indication that the signal 
is intrinsic to the star.
This also could explain 
why no transits were observed by {\it K2}
\citep{Cloutier:2017},
although this can also be explained by misaligned orbits.


\section{Joint HARPS and CARMENES Analysis} \label{sec:both-sets}

\begin{figure}[t!]
	\includegraphics[width=0.5\textwidth]{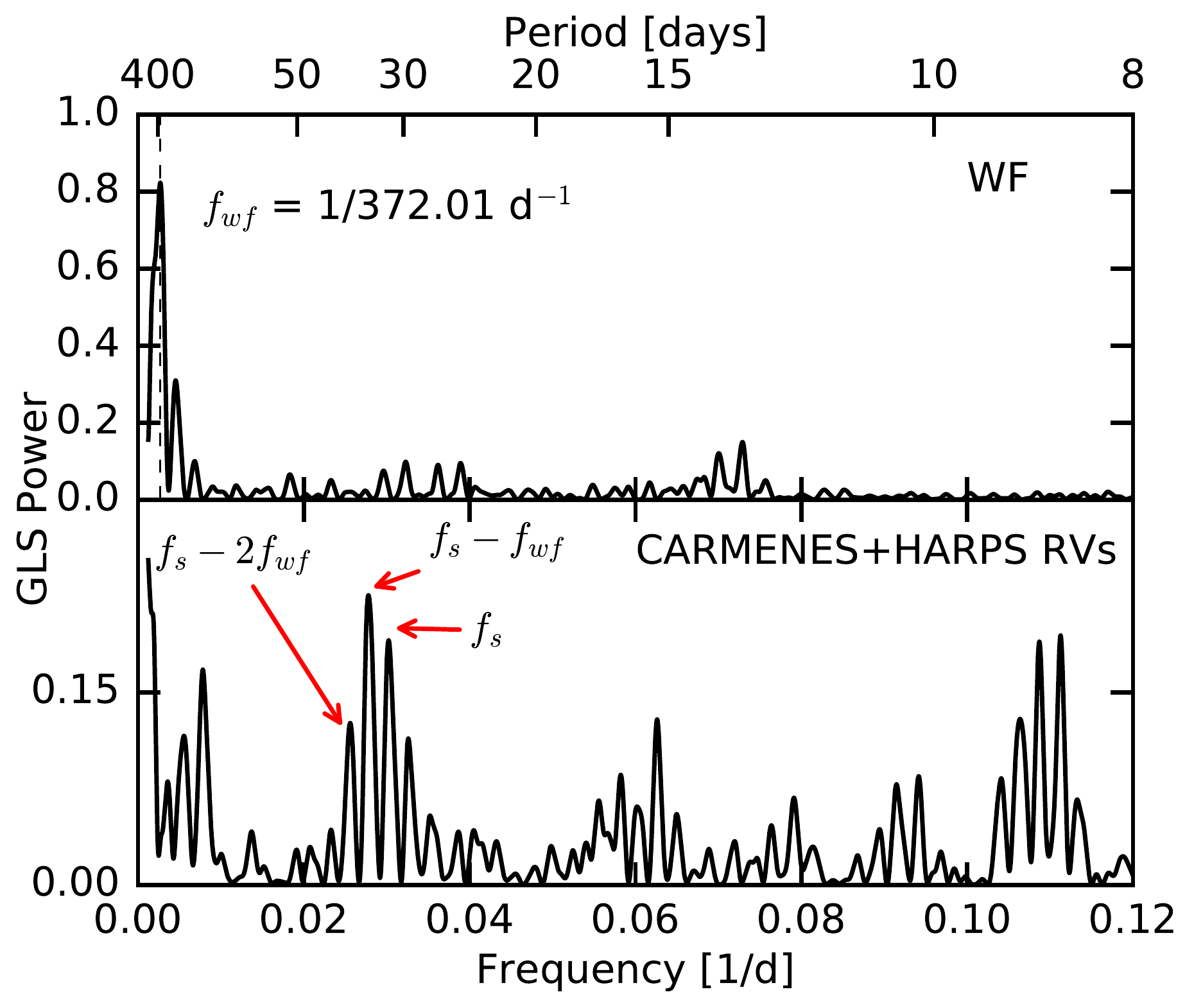}
	\caption{Periodogram of the WF (top) and RVs for the combined HARPS and CARMENES data set (bottom). The WF shows a significant peak at the sidereal year. The aliases of the planetary signal are indicated by the red arrows. 
	\label{fig:combined-periodo}}
\end{figure}

In this Section, we combine
both the HARPS and CARMENES-VIS data sets 
to refine the parameters of the system,
in particular to put constraints on
the eccentricity. 
The joint HARPS (75 observations) and 
CARMENES-VIS (58 observations) data sets contain 
a total of 133 RV measurements 
with a time baseline of 807 days.
A periodogram analysis for the WF of the combined set
reveals a peak at $\sim$ 372 days
(Figure~\ref{fig:combined-periodo}, upper panel).
This is expected since the data set spans three seasons 
with gaps in between.
However, if there is a signal in the raw RVs at frequency $f_s$,
then in the periodogram, aliases will likely appear at 
$f_\mathrm{alias,n}$ = $f_s + nf_\mathrm{WF}$,
where $n$ is an integer and $f_\mathrm{WF}$ is the frequency 
at which the WF shows a peak 
(also known as the sampling frequency)
\citep{Dawson:2010}.
Considering that the RV signal due to the 
transiting planet is present in the data, 
then
$f_\mathrm{alias,1} = 1/32.9396 - 1/372.01 = 0.02767$ day$^{-1}$
($\sim$ 36.14 days).
For $n = 2$, $f_\mathrm{alias,2} = 0.02498$ day$^{-1}$ 
($\sim$ 40.03 days).
This means that an alias of the orbital frequency of the planet 
is right at the stellar rotation frequency.
Similarly, the aliases of the stellar rotation frequency 
are also approximately at 33 and 36 days. 
It is a coincident that the alias of one signal is 
close to the real frequency of the other signal.
It is also by chance that the aliases of both signals meet 
at 36 days. 
So these aliases interfere and give a higher GLS power at this frequency.
The aliases are shown 
in the lower panel of Figure~\ref{fig:combined-periodo}.

We performed a Keplerian fit for the combined HARPS 
and CARMENES-VIS RVs
using the publicly available python package
\texttt{RadVel}\footnote{
\url{https://github.com/California-Planet-Search/radvel}}
\citep{fulton:2018}.
\texttt{RadVel} is capable of modeling RV data
taken with different instruments and
uses a fast Keplerian equation solver written in C
and the \texttt{emcee} ensemble sampler \citep{dfm:2013}.
The optical fibres of the HARPS spectrograph
were upgraded in June 2015 \citep{LoCurto:2015}.
Consequently, this affected the radial velocity offset
and therefore we treated the data taken
pre- and post-fibre upgrade separately
by accounting a different velocity offset for 
each data set 
($\gamma_\mathrm{preHARPS}$ and $\gamma_\mathrm{postHARPS}$).
We account for stellar activity by adding an RV jitter term.
Three independent jitter terms 
($\sigma_\mathrm{preHARPS}$, $\sigma_\mathrm{postHARPS}$, 
$\sigma_\mathrm{CARMENES}$)
were added in quadrature to the formal error bars of each instrument
and were allowed to vary.
We followed \cite{Ford:2005}
and fit for
$\sqrt{e}\cos\omega$ and $\sqrt{e}\sin\omega$
instead of the eccentricity $e$ and 
argument of periastron $\omega$ 
to increase the rate of convergence.
We thus fit for eleven parameters:
the planetary semi-amplitude $K_b$, 
$\sqrt{e}\cos\omega$,
$\sqrt{e}\sin\omega$,
planetary orbital period $P_b$,
time of conjunction $T_c$,
the velocity offsets for the CARMENES, HARPS pre-fibre,
and HARPS post-fibre upgrade, $\gamma_\mathrm{CARMENES}$, 
$\gamma_\mathrm{preHARPS}$, and $\gamma_\mathrm{postHARPS}$,
and for 
$\sigma_\mathrm{preHARPS}$, $\sigma_\mathrm{postHARPS}$, 
and $\sigma_\mathrm{CARMENES}$.
We assign Gaussian priors on $P_b$ and $T_c$,
adopt uniform uninformative priors on the jitter and offset terms
and measure $e = 0.20 \pm 0.08$
and $K_b = 3.55^{+0.57}_{-0.58}$ \mpersec.
This translates into a planetary mass 
$M_\mathrm{b} = 8.92^{+1.70}_{-1.60} \, M_{\oplus}$,
consistent with the previous analysis using
only the CARMENES-VIS data set (Section \ref{sec:orbital-analysis}).
The median values and the 68\% credible intervals
are reported in Table \ref{tab:combinedanalysis}.
The joint and marginalized posterior
constraints on the model parameters
are shown in Figure~\ref{fig:comb-corner-plot} 
and Figure~\ref{fig:ecc-dist} shows the eccentricity distribution.

\begin{table}[t!]
\caption{Orbital and planetary parameters for the system \ktwo\ for the combined HARPS and CARMENES-VIS data sets
\label{tab:combinedanalysis}}
\begin{center}
\begin{tabular}{ll}
\toprule
Parameter & Value \\
\hline
Orbital Parameters & \\
$ \qquad T_0 $ [BJD]                & $2457264.39144 \pm 0.00065$  \\
$ \qquad P_b $ [d]                  & $32.939623^{+0.000095}_{-0.000100}$ \\
$ \qquad K_b \left[\mpersec\right]$ & $3.55^{+0.57}_{-0.58}$ \\
$ \qquad e_b$                       & $0.20 \pm 0.08$ \\
$ \qquad \omega_b$ [rad]            & $-0.10^{+0.81}_{-0.59}$ \\
\hline
Planetary Parameters & \\
$ \qquad R_\mathrm{p}$ [$R_{\oplus}$] \tablenotemark{b}  & 2.37$ \pm 0.22$           \\
$ \qquad i $ [deg]                    \tablenotemark{a}  & 89.5785$^{+0.0079}_{-0.0088}$       \\
$ \qquad a $ [au]                     \tablenotemark{a}  & 0.1429$^{+0.006}_{-0.0065}$         \\
$ \qquad M_\mathrm{b}$ [$M_{\oplus}$]         & $8.92^{+1.70}_{-1.60}$ \\
$ \qquad T_{\mathrm{eq,b}}$ [K]                 & $284 \pm 15$\\
$ \qquad \rho_{\mathrm{b}}$ [g cm$^{-3}$]     & $4.11^{+1.72}_{-1.18}$   \\
\hline
Other Parameters & \\
$ \qquad \gamma_\mathrm{CARMENES} \left[\mpersec\right]$  & $-3.40 \pm 0.56$ \\
$ \qquad \gamma_\mathrm{preHARPS} \left[\mpersec\right]$  & $656.4^{+1.8}_{-1.9}$ \\
$ \qquad \gamma_\mathrm{postHARPS} \left[\mpersec\right]$ & $653.86^{+0.61}_{-0.59}$ \\
$ \qquad \sigma_\mathrm{CARMENES} \left[\mpersec\right]$  & $2.78^{+0.61}_{-0.53}$ \\
$ \qquad \sigma_\mathrm{preHARPS} \left[\mpersec\right]$  & $2.5^{+2.5}_{-1.7}$ \\
$ \qquad \sigma_\mathrm{postHARPS} \left[\mpersec\right]$  & $3.06^{+0.69}_{-0.64}$ \\
\hline
\end{tabular}
\tablenotetext{a}{Parameters based on \cite{Benneke:2017}.}
\tablenotetext{b}{Recalculated the value using $R_\mathrm{b}/R_*$ and $R_*$ as derived by \cite{Benneke:2017}.}
\end{center}
\end{table}

\section{Discussion} \label{sec:results}

Using the CARMENES-VIS data only, 
we detected \ktwo\ with a semi-amplitude of 
$K = 3.60_{-0.51}^{+0.53}$ \mpersec,
in agreement with the value estimated by 
\cite{Cloutier:2017} using data taken with HARPS.
We then combined the CARMENES-VIS and HARPS data sets 
to refine the planetary parameters,
particularly to put constraints on the eccentricity.
We derived a semi-amplitude of 
$K_b = 3.55^{+0.57}_{-0.58}$ \mpersec\
and eccentricity $e = 0.20 \pm 0.08$
indicating that the planet is on a slightly eccentric orbit.
This implies a mass 
$M_\mathrm{b} = 8.92^{+1.70}_{-1.60} \, M_{\oplus}$ that,
combined
with the radius estimate we derived in Section~\ref{sec:orbital-analysis} 
$R_\mathrm{b} = 2.37 \pm 0.22 \, R_{\oplus}$,
leads to a bulk density of 
$\rho_b = 4.18^{+1.71}_{-1.17} \, \mathrm{g \,cm^{-3}}$.
However, the radius estimate could be affected 
by systematic errors due to stellar contamination
\citep{Rackham:2018}.
Consequently, this leads to
systematic errors in the derived density.

\begin{figure}[t]
\begin{center}
	\includegraphics[width=1\columnwidth]{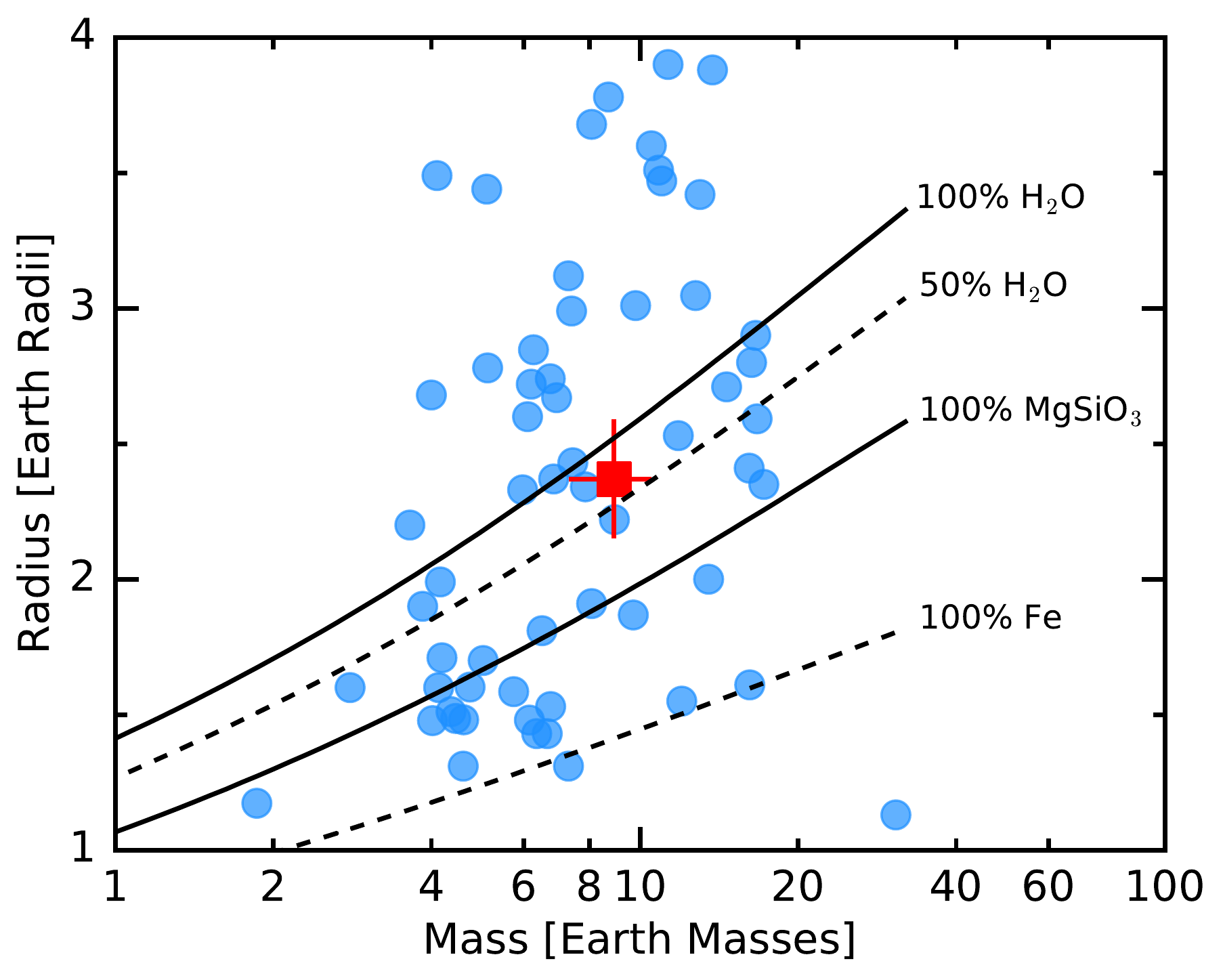}
	\caption{ Mass-radius diagram for well characterized transiting exoplanets. \ktwo\ (red square) and theroretical models \citep{Zeng:2016} are overplotted. The composition of the planet is consistent with 50\% H$_2$0 and 50\% MgSiO$_3$.} 
	\label{fig:mrdiagram}
\end{center}
\end{figure}

We put the parameters of \ktwo\
in the context of discovered exoplanets 
of similar sizes and masses.
Figure~\ref{fig:mrdiagram} shows
the position of \ktwo\ on the mass-radius diagram
in comparison with the other discovered exoplanets
\footnote{Data taken on the 6$^\mathrm{th}$ of November from NASA Exoplanet Archive 
\url{http://exoplanetarchive.ipac.caltech.edu}} 
with radii less than $4 R_{\oplus}$,
masses smaller than 32 $M_{\oplus}$,
and with masses and radii determined 
with a precision better than 30\%.
Theoretical two-layer models
obtained from \cite{Zeng:2016} are overplotted.
It can be seen that \ktwo\ can have a composition consistent with 
$\sim$ 100\% water (H$_2$O) or
$\sim$ 50\% H$_2$O and $\sim$ 50\% rock (MgSiO$_3$)
indicating that this planet could be water rich.
However, it is well known
that there is a wide range of possible compositions
for a given mass and radius,
all of which include low density volatiles such as water and H/He
\citep{Lopez:2012,Jin:2018}.
The radius of \ktwo\
can be thus explained by a silicate and iron core 
along with a H/He envelope or with a water envelope.
This is in agreement with \cite{Rogers:2015} and \cite{Wolfgang:2015},
who showed that most planets with radii larger than 1.6 R$_\oplus$ are not rocky.

Transiting low-mass planets in the temperate zone of M stars 
are potential prime targets for detailed atmospheric characterization.
\ktwo\ lies in the temperate zone of its host star
\citep{Kopparapu:2013,Kopparapu:2014}
and receives stellar irradiation similar to Earth.
In addition to that, 
the brightness of the star in the NIR
($J = 9.8$ mag and $K = 8.9$ mag)
and its close distance 
makes \ktwo\
a good candidate
for detailed atmospheric characterization
with observations of secondary transits.
The {\it James Webb Space Telescope}
will be able to simultaneously observe 
from 0.6 to 2.8 $\mu$m
and thus can provide robust detections 
of water absorption bands in the NIR (if any)
for this bright target.

\section{Conclusions} \label{sec:conclusion}

\ktwo\ was first discovered as part of the {\it K2} mission
\citep{Montet:2015}.
Later, \cite{Benneke:2017} confirmed the presence of the planet
by detecting a third transit light curve of the same depth using {\it Spitzer}.
We obtained contemporaneous photometric 
and spectroscopic observations to 
model jointly stellar activity and 
the Keplerian signal of \ktwo.
We found the stellar rotation period $P_\mathrm{rot}$
to be close to the planetary orbital period,
in agreement with {\it K2} photometry
\citep{Cloutier:2017,Stelzer:2016}.
The simultaneous photometric data
along with the precise RV observations
were a key to disentangle these two signals.
Coincidentally, the window function 
also shows a peak close to the orbital 
period of the planet. 
We performed several tests to assess whether 
the RV signal due to the planet is detected in the RV data
and to test whether stellar activity affects the determination 
of the planetary amplitude. 
Our analysis highlights the difficulty 
in detecting non-transiting low-mass planets 
in the presence of uneven sampling 
and, more importantly, when 
the planetary signal is close to the stellar rotation period.

Using data taken with HARPS,
\cite{Cloutier:2017} claimed that the system
hosts two planets:
($i$) an outer planet, \ktwo, with an amplitude of 
$K_b = 3.18 \pm 0.71$ \mpersec,
($ii$) an inner non-transiting planet, \ktwoc, 
which has a higher signal compared to \ktwo,
and a period of $8.962 \pm 0.008$ days.
While the existence of \ktwo\ 
is in agreement with results derived with the CARMENES-VIS data,
the 9-day signal in our data set
is not significant
and only present in the blue part of the spectrum 
when the star is showing high activity levels.
We thus believe that the signal is time and color variable,
and is correlated with the chromospheric stellar activity.
K2-18\,c is mostly an artifact of stellar activity 
and not a bona fide planet.
This analysis underscores the importance of
multi-wavelength RV observations, 
in particular the value of comparing 
the blue and red orders of active stars
to check the consistency of planetary signals
across all orders of the Echelle spectrum.

Disentangling the signal of a low-mass planet
from the stellar RV signal is still challenging.
Following \cite{Vanderburg:2016},
we also encourage 
future studies to perform a 
combined analysis of simultaneous photometry, 
multi-wavelength RV observations,
and analysis of the activity indicators 
to overcome these challenges and 
to test the reliability of signals 
present in the data.


\acknowledgments

CARMENES is an instrument for the Centro Astron\'omico Hispano-Alem\'an de
  Calar Alto (CAHA, Almer\'{\i}a, Spain). 
  CARMENES is funded by the German Max-Planck-Gesellschaft (MPG), 
  the Spanish Consejo Superior de Investigaciones Cient\'{\i}ficas (CSIC),
  the European Union through FEDER/ERF FICTS-2011-02 funds, 
  and the members of the CARMENES Consortium 
  (Max-Planck-Institut f\"ur Astronomie,
  Instituto de Astrof\'{\i}sica de Andaluc\'{\i}a,
  Landessternwarte K\"onigstuhl,
  Institut de Ci\`encies de l'Espai,
  Insitut f\"ur Astrophysik G\"ottingen,
  Universidad Complutense de Madrid,
  Th\"uringer Landessternwarte Tautenburg,
  Instituto de Astrof\'{\i}sica de Canarias,
  Hamburger Sternwarte,
  Centro de Astrobiolog\'{\i}a and
  Centro Astron\'omico Hispano-Alem\'an), 
  with additional contributions by the Spanish Ministry of Economy, 
  the German Science Foundation (DFG), 
  the states of Baden-W\"urttemberg and Niedersachsen, 
  and by the Junta de Andaluc\'{\i}a.
P.S. would like to thank Christoph Mordasini for his helpful comments on an early draft and
Sudeshna Boro-Saikia for her help on calculating the calcium indices of the HARPS data set.
I.R. and J.C.M. acknowledge support by the Spanish Ministry of Economy and Competitiveness (MINECO) and the Fondo Europeo de Desarrollo Regional (FEDER) through grant ESP2016-80435-C2-1-R, as well as the support of the Generalitat de Catalunya/CERCA programme.
APH acknowledges the support of the Deutsche Forschungsgemeinschaft (DFG) grant HA 3279/11-1.
J.A.C., P.J.A and D.M. acknowledge support by the Spanish Ministry of Economy and Competitiveness (MINECO) from projects AYA2016-79425- C3-1,2,3-P.
VJSB is supported by programme AYA2015-69350-C3-2-P 
from Spanish Ministry of Economy and Competitiveness (MINECO).
We thank the anonymous referee for useful comments
that improved the paper.


\software{astropy \citep{Astropy:2013},
		  corner \citep{dfm:2016},
          emcee \cite{dfm:2013},
          matplotlib \citep{Hunter:2007},
          RadVel \citep{fulton:2018}}




\label{bib}
\bibliographystyle{aasjournal}
\bibliography{k218.bib} 



\newpage

\appendix 

\section{Additional Figures and Tables} 
\label{sec:appendix}

\begin{figure*}
\begin{minipage}{0.24\textwidth}
	 \centering
        \includegraphics[width=1\textwidth]{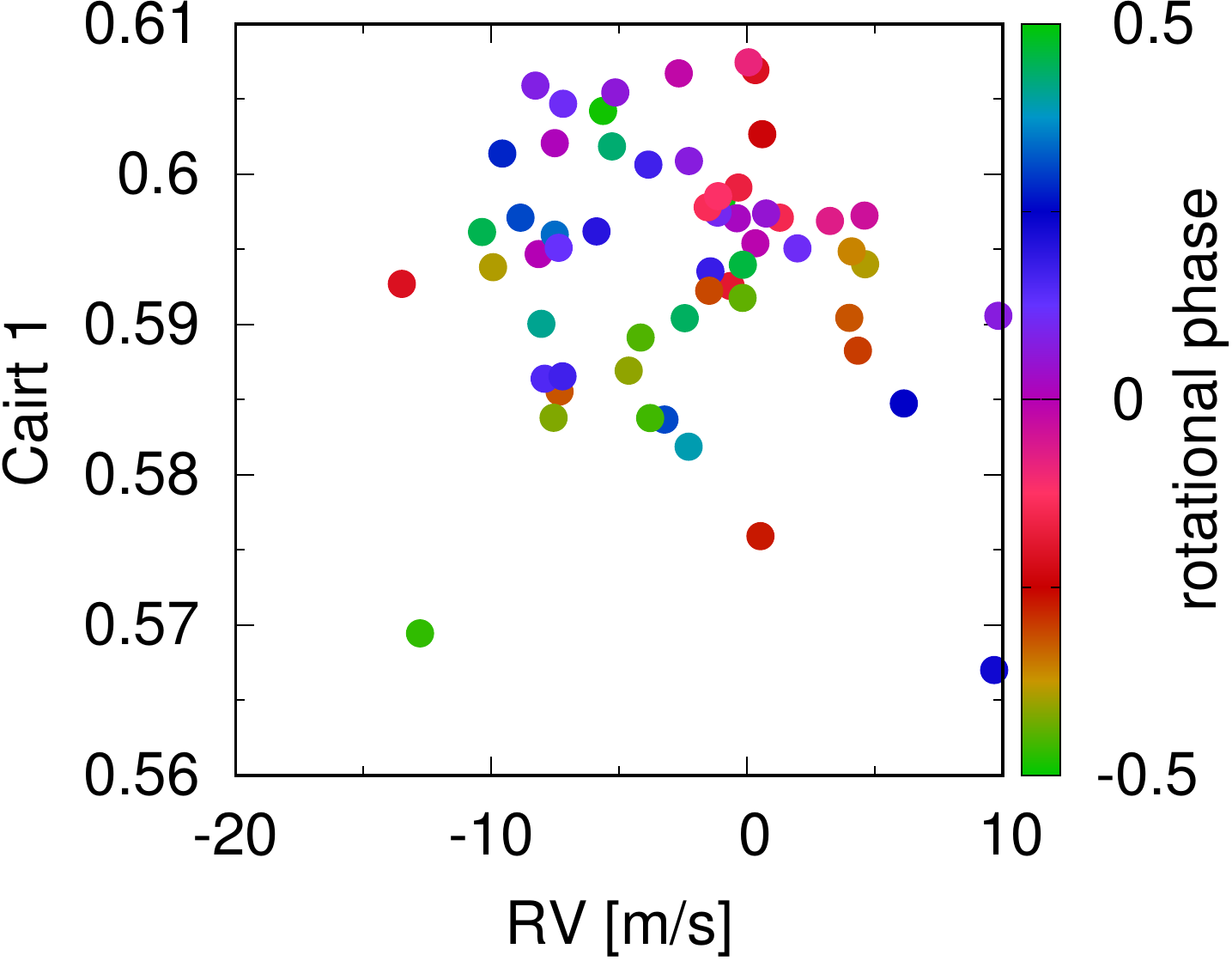}
        \includegraphics[width=1\textwidth]{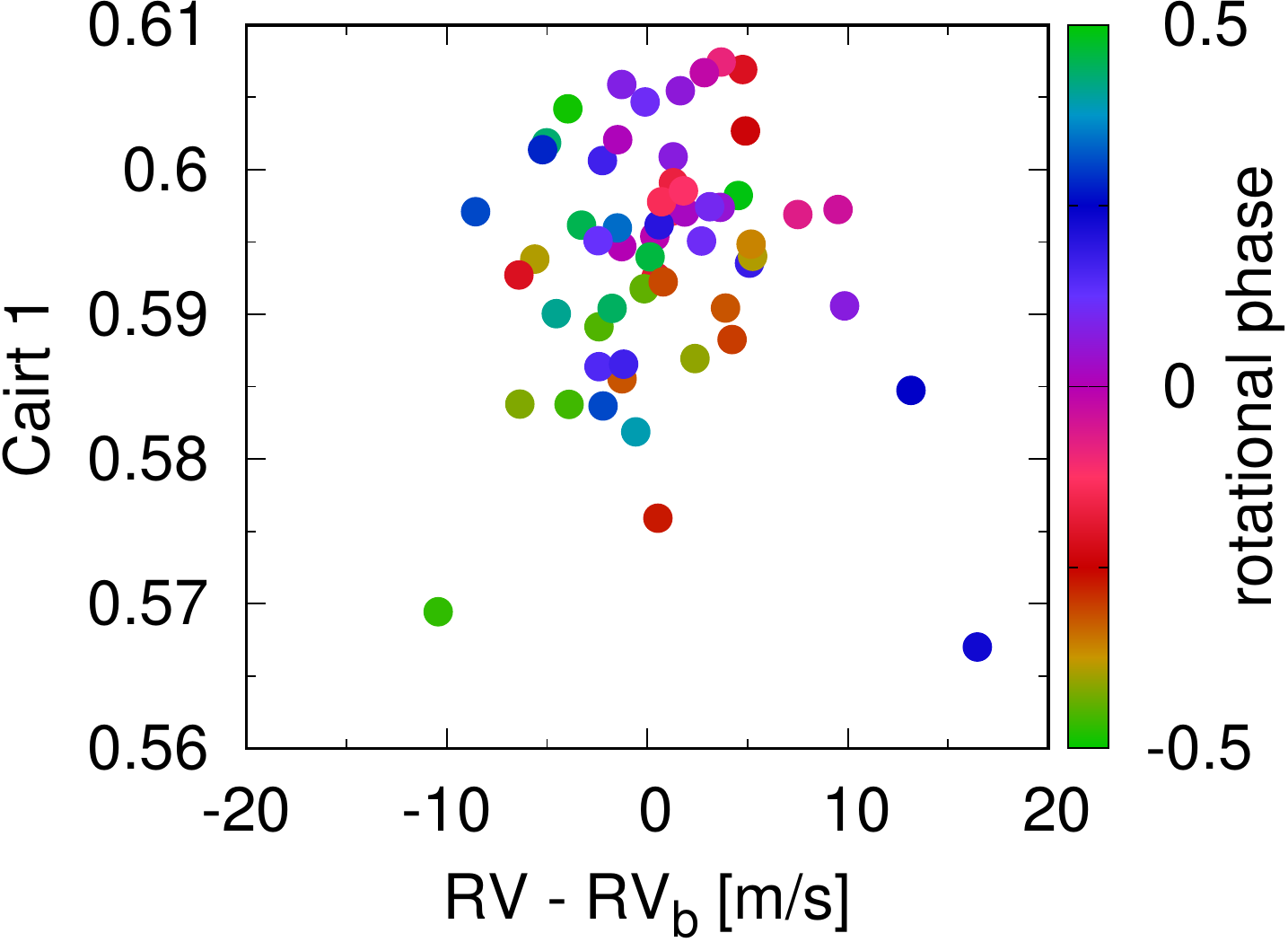}

     \end{minipage}
     \begin{minipage}{0.24\textwidth}
      \centering
       \includegraphics[width=1\textwidth]{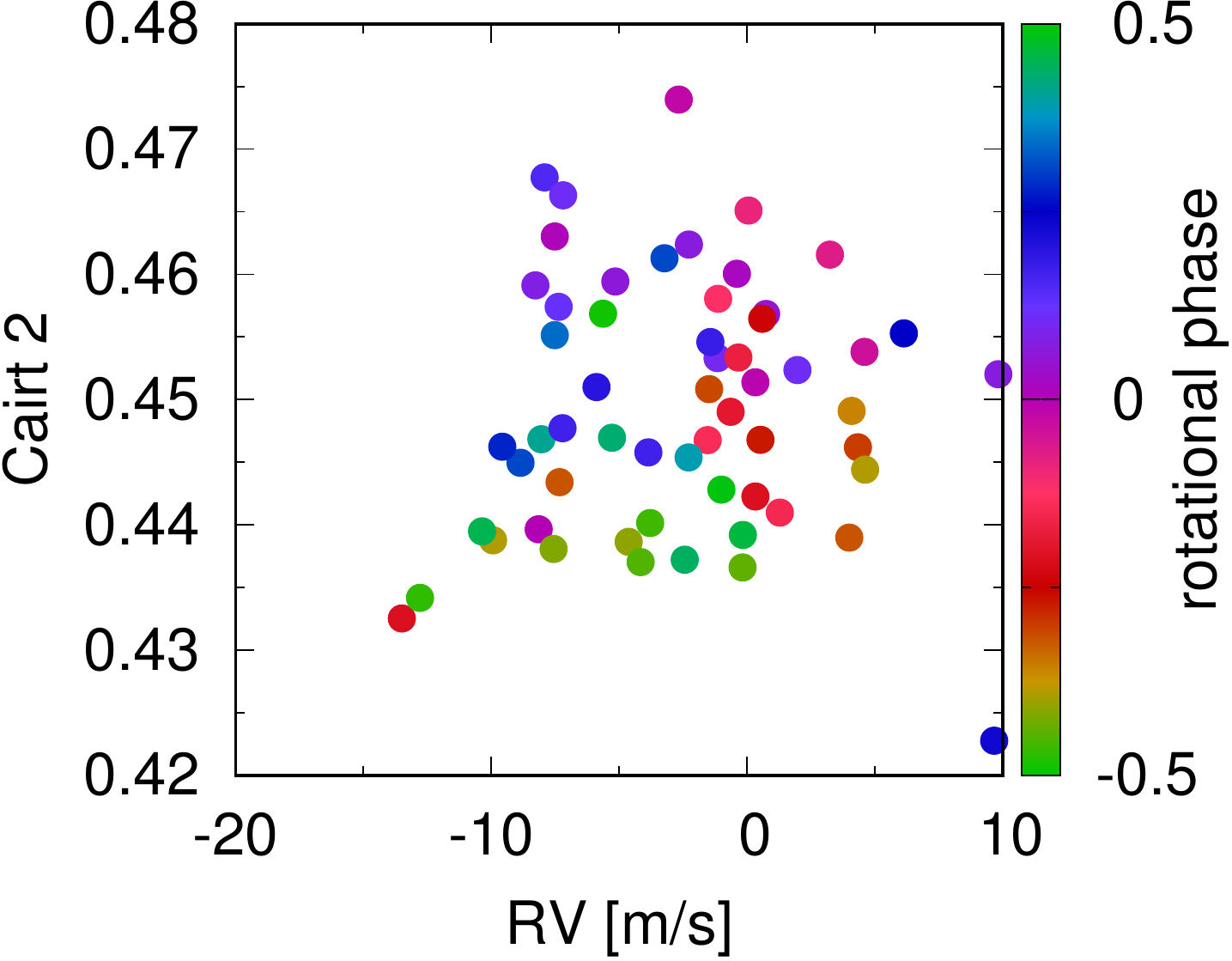}
       \includegraphics[width=1\textwidth]{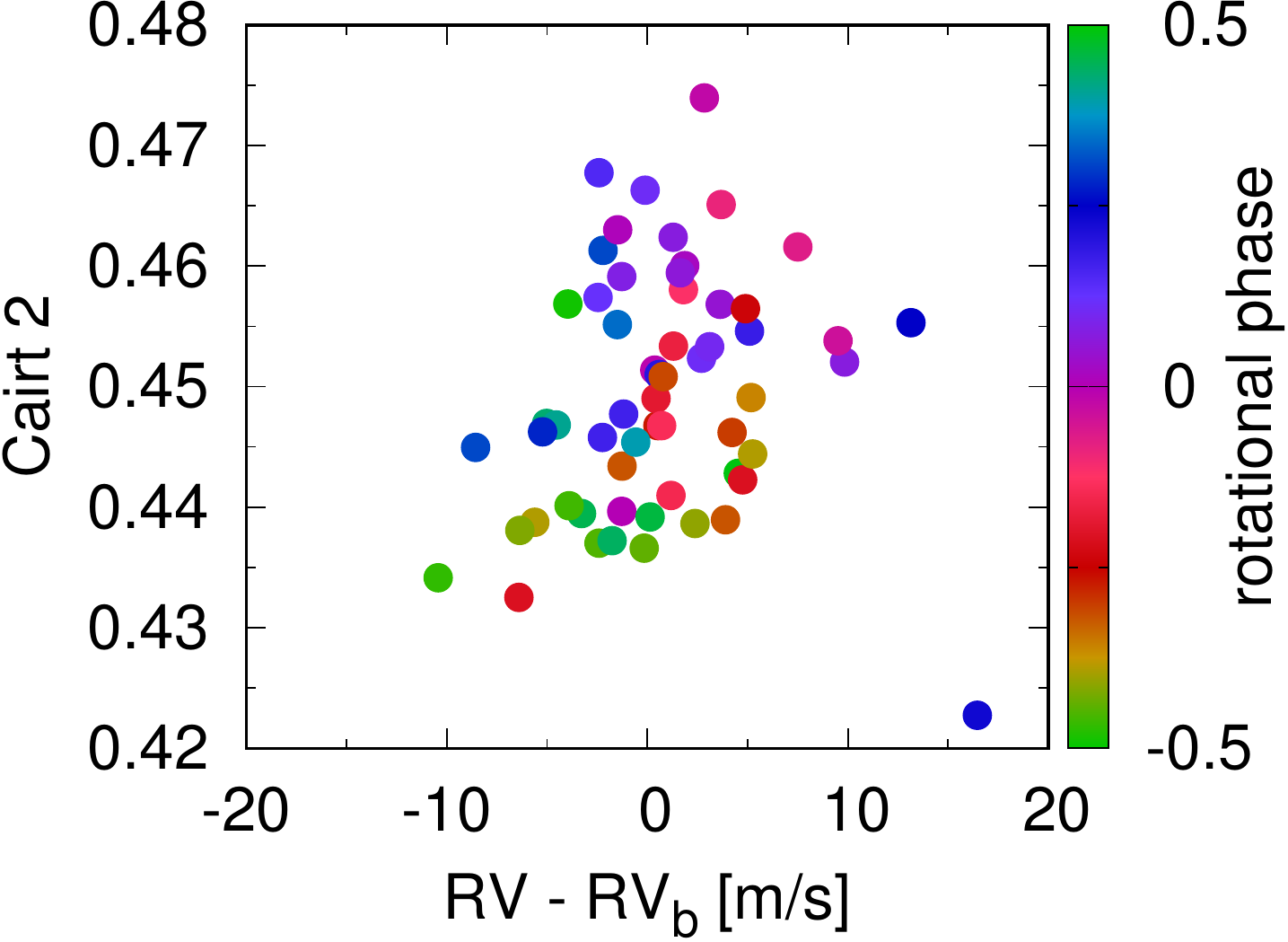}
       
     \end{minipage}
     \begin{minipage}{0.24\textwidth}
      \centering
        \includegraphics[width=1\textwidth]{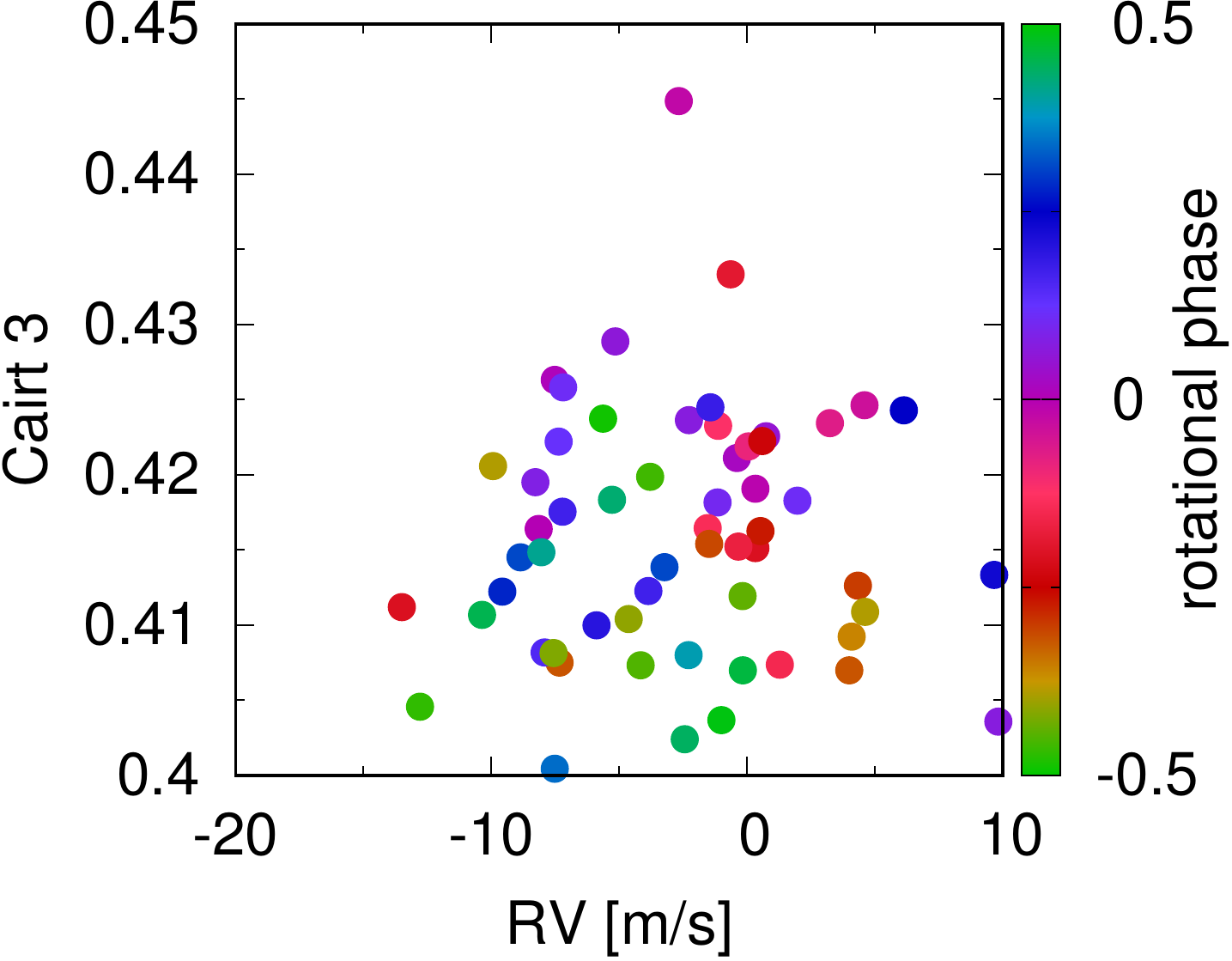}
        \includegraphics[width=1\textwidth]{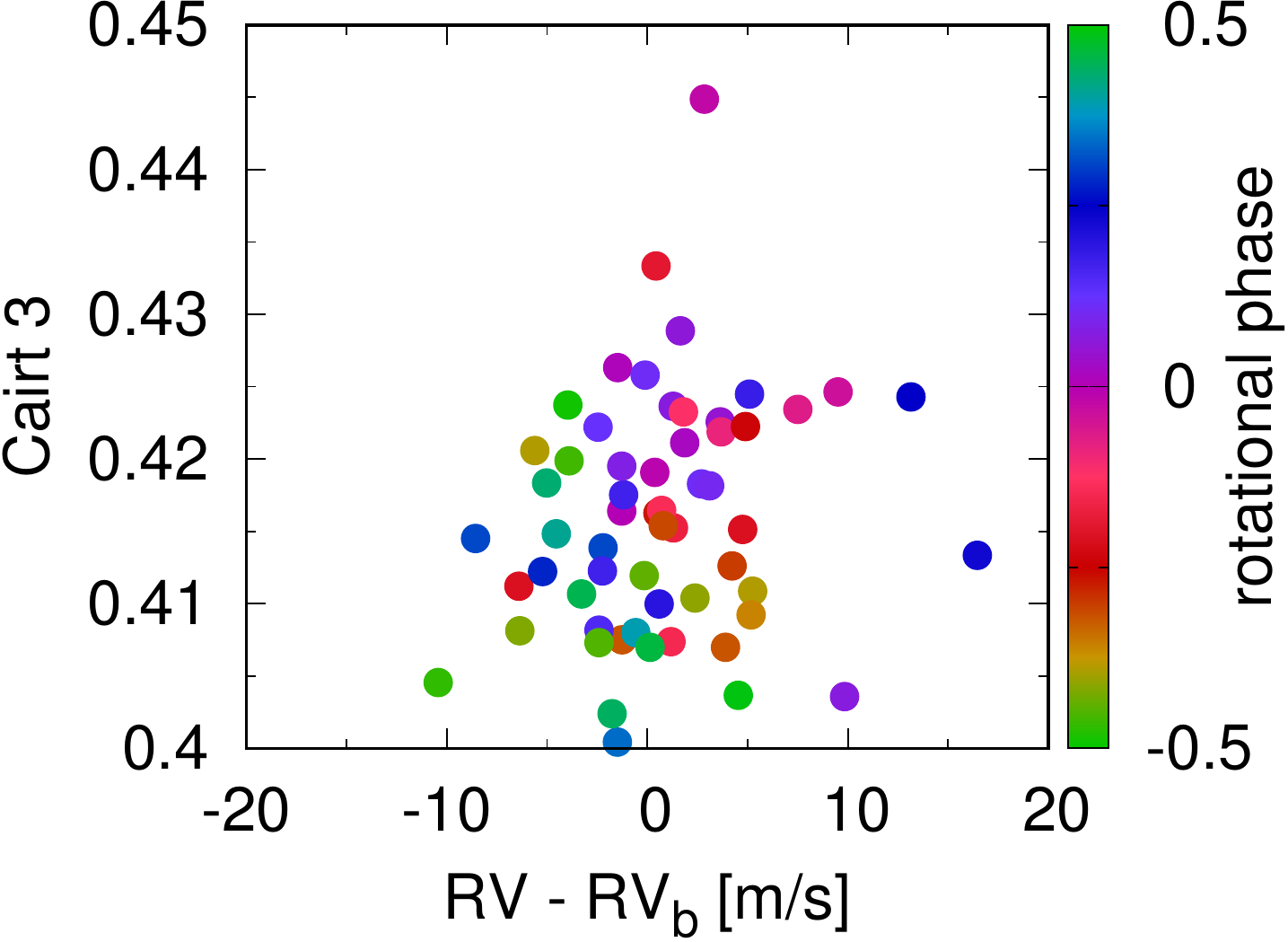}

     \end{minipage}
     \begin{minipage}{0.24\textwidth}
      \centering
        \includegraphics[width=1\textwidth]{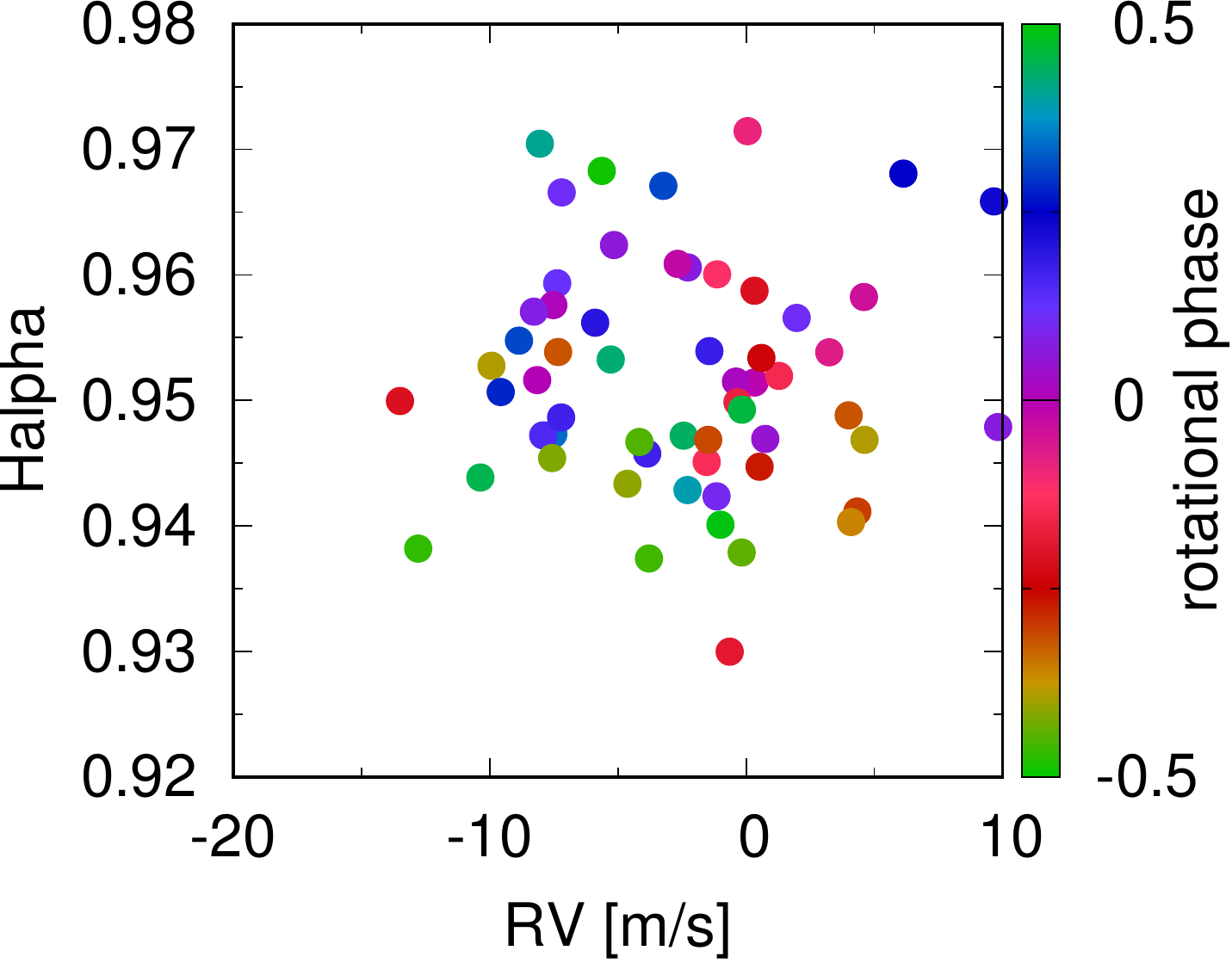}
        \includegraphics[width=1\textwidth]{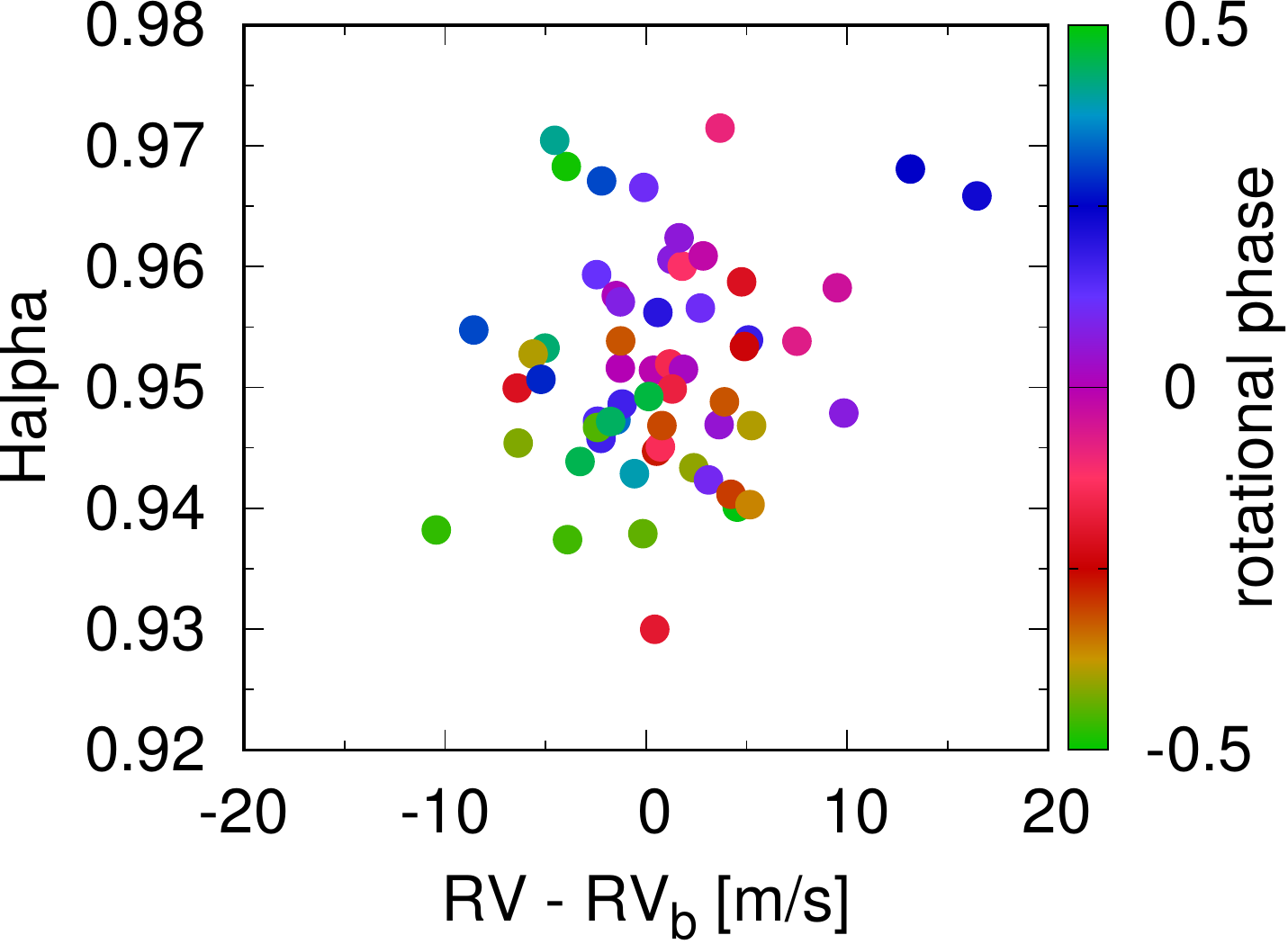}

     \end{minipage}
\caption{Upper panels: measured RVs plotted against various stellar activity indicators phase folded using the stellar rotation period of 39.63 days. Lower panels: same as the upper panels but after subtracting the planetary signal. None of the activity indicators show a statistically significant linear or circular correlation with the raw RVs or the residuals.}
	\label{fig:rv-act-corr}
\end{figure*}

\begin{figure}[t!]
	\includegraphics[width=\textwidth]{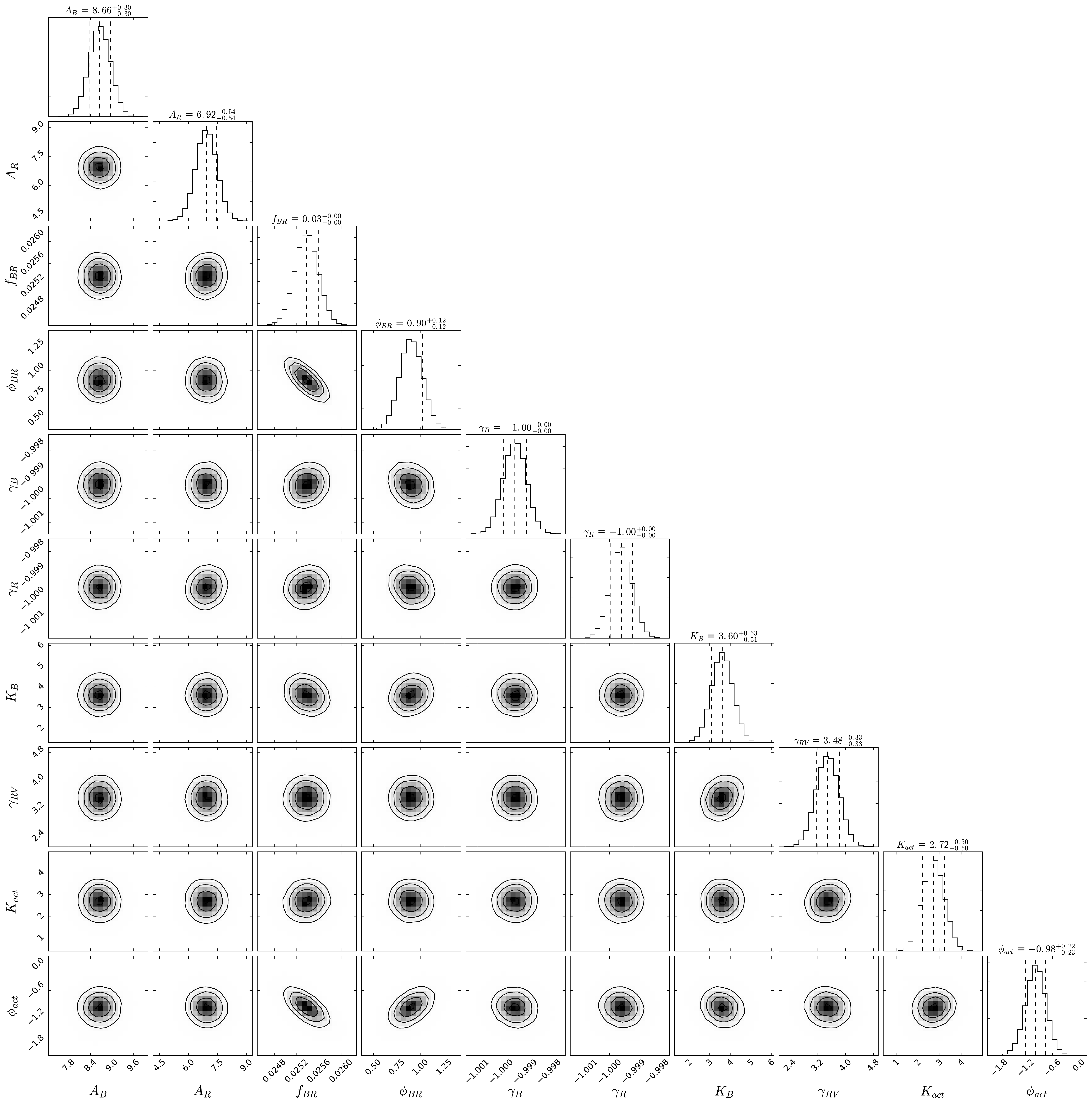}
	\caption{The marginalized posterior distributions on the model parameters from the joint analysis of the photometry and CARMENES RV measurements.}
	\label{fig:carmenes-corner-plot}
\end{figure}

\begin{figure}[t!]
	\includegraphics[width=\textwidth]{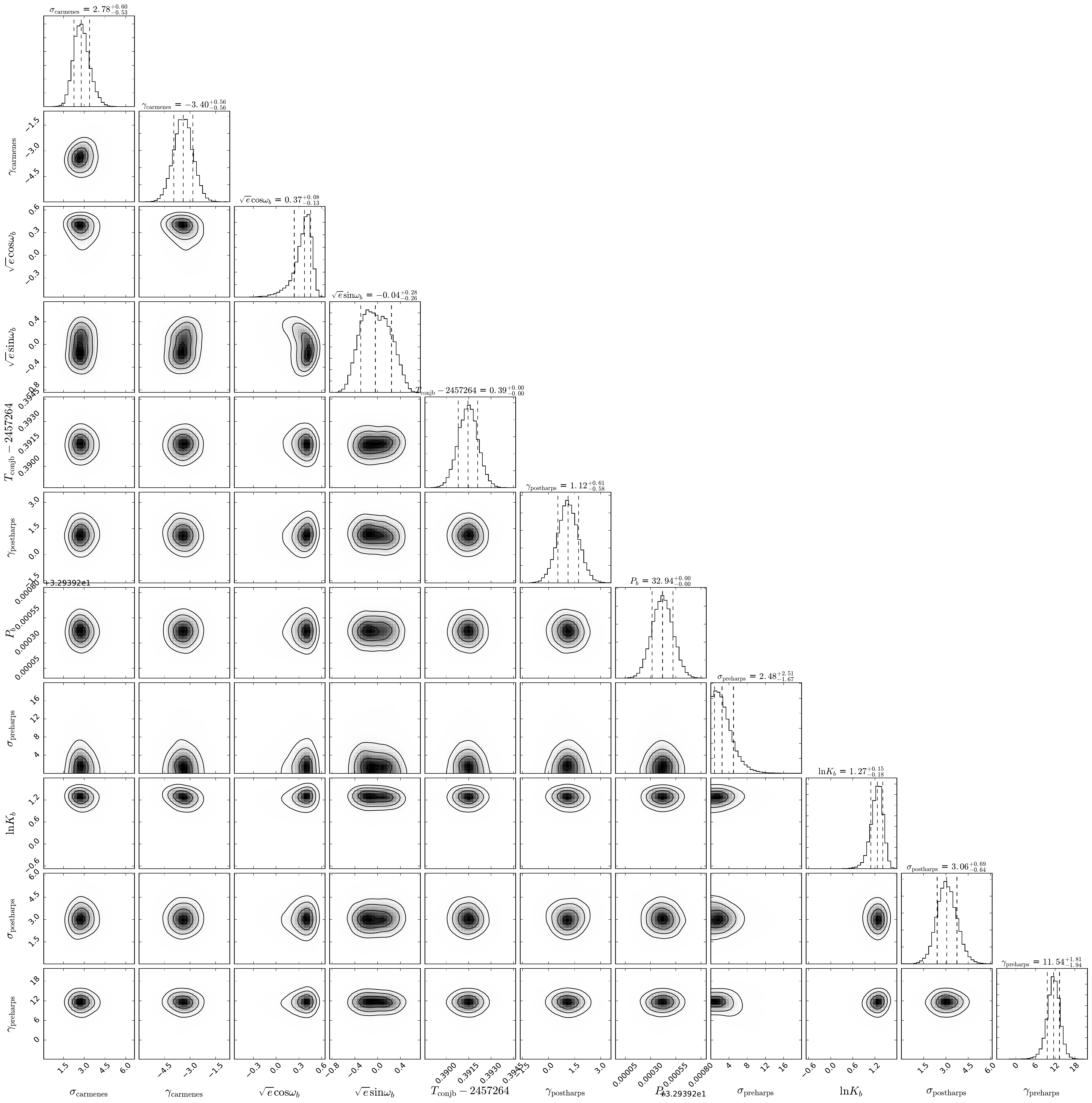}
	\caption{The marginalized posterior distributions on the model parameters of the RV measurements using CARMENES and HARPS data.}
	\label{fig:comb-corner-plot}
\end{figure}

\begin{figure}[t!]
	\includegraphics[width=0.5\textwidth]{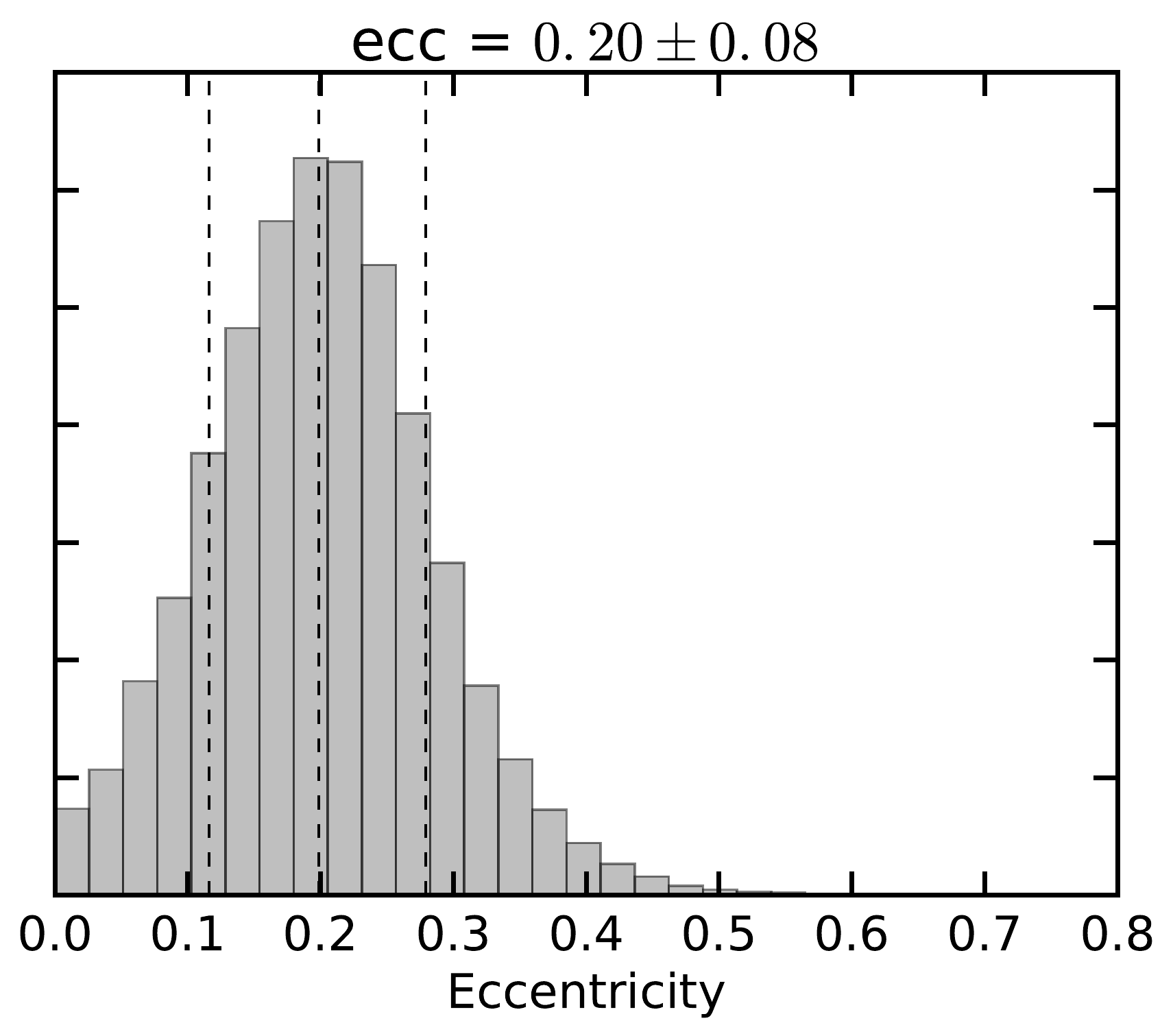}
	\caption{The posterior distribution of $e$ calculated from $\sqrt{e} \cos \omega_b$ and $\sqrt{e} \sin \omega_b$. The vertical lines show the 16$^\mathrm{th}$, 50$^\mathrm{th}$, and 84$^\mathrm{th}$ percentile.}
	\label{fig:ecc-dist}
\end{figure}

\begin{table}[h]
\caption{WiFSIP photometry in the Johnson {\it B} filter.}
\label{tab:data-photb}
\begin{center}
\begin{tabular}{ccc}
\hline
\hline
BJD $-2450000$ &  $\Delta B$  & $\sigma_B$ \\
(days) & (mag) & (mag) \\
\hline
7812.628906 & 0.9934 & 0.0023 \\
7813.632812 & 0.9894 & 0.0020 \\
7815.636719 & 0.9928 & 0.0022 \\
7816.625000 & 0.9889 & 0.0028 \\
7817.597656 & 0.9896 & 0.0062 \\
7818.628906 & 0.9965 & 0.0024 \\
7819.585938 & 0.9886 & 0.0052 \\
7833.562500 & 1.0046 & 0.0022 \\
7834.570312 & 1.0053 & 0.0027 \\
7836.550781 & 1.0098 & 0.0022 \\
7838.546875 & 1.0079 & 0.0019 \\
7841.531250 & 1.0052 & 0.0023 \\
7842.546875 & 1.0026 & 0.0021 \\
7843.546875 & 1.0039 & 0.0023 \\
7846.515625 & 1.0039 & 0.0045 \\
7856.492188 & 0.9930 & 0.0032 \\
7858.097656 & 0.9880 & 0.0044 \\
7860.515625 & 0.9962 & 0.0024 \\
7874.417969 & 1.0031 & 0.0020 \\
7875.398438 & 1.0052 & 0.0021 \\
7892.378906 & 0.9920 & 0.0022 \\
7897.390625 & 0.9855 & 0.0036 \\
7901.390625 & 0.9865 & 0.0022 \\
7910.410156 & 1.0153 & 0.0027 \\
7913.429688 & 1.0142 & 0.0032 \\
7916.386719 & 1.0095 & 0.0028 \\
7921.402344 & 1.0102 & 0.0033 \\
\hline
\end{tabular}
\end{center}
\end{table}

\begin{table}[h]
\caption{WiFSIP photometry in the Cousins {\it R} filter.}
\label{tab:data-photr}
\begin{center}
\begin{tabular}{ccc}
\hline
\hline
BJD $-2450000$ &  $\Delta R$   & $\sigma_R$ \\
(days) & (mag) & (mag) \\
\hline
7812.628906 & 0.9929 & 0.0018 \\
7813.632812 & 0.9904 & 0.0018 \\
7815.636719 & 0.9937 & 0.0019 \\
7816.628906 & 0.9920 & 0.0049 \\
7817.601562 & 0.9928 & 0.0026 \\
7818.628906 & 0.9977 & 0.0023 \\
7819.585938 & 0.9921 & 0.0027 \\
7833.562500 & 1.0026 & 0.0021 \\
7834.570312 & 1.0014 & 0.0023 \\
7836.554688 & 1.0115 & 0.0023 \\
7838.546875 & 1.0052 & 0.0020 \\
7841.531250 & 1.0040 & 0.0021 \\
7842.546875 & 1.0013 & 0.0024 \\
7843.550781 & 1.0049 & 0.0054 \\
7846.515625 & 1.0030 & 0.0030 \\
7856.496094 & 1.0001 & 0.0057 \\
7857.753906 & 0.9932 & 0.0045 \\
7860.515625 & 0.9970 & 0.0036 \\
7874.417969 & 1.0057 & 0.0021 \\
7875.398438 & 1.0094 & 0.0020 \\
7892.378906 & 0.9967 & 0.0034 \\
7897.394531 & 0.9970 & 0.0036 \\
7901.390625 & 0.9917 & 0.0021 \\
7910.414062 & 1.0093 & 0.0025 \\
7913.433594 & 1.0081 & 0.0020 \\
7916.386719 & 1.0055 & 0.0028 \\
7921.406250 & 1.0012 & 0.0021 \\
\hline
\end{tabular}
\end{center}
\end{table}

\clearpage

\begin{longrotatetable}
\begin{deluxetable*}{ccccccccccccccc}
\tablecaption{ Radial Velocities obtained with CARMENES and the spectroscopic activity indicators
\label{tab:data-carmenes}}
\tabletypesize{\scriptsize}
\tablehead{
\colhead{BJD $- 2450000$} & \colhead{RV} & \colhead{$\sigma_{RV}$} & \colhead{blue RV} & \colhead{$\sigma_\mathrm{blue}$} & \colhead{red RV} & \colhead{$\sigma_\mathrm{red}$} & {Ca\,{\sc ii}~IRT~1} & {$\sigma_\mathrm{{CaIRT1}}$} & {Ca\,{\sc ii}~IRT~2} & {$\sigma_\mathrm{{CaIRT2}}$} & {Ca\,{\sc ii}~IRT~3} & {$\sigma_\mathrm{{CaIRT3}}$} & {H$\alpha$} & {$\sigma_{H_\alpha}$} \\
\colhead{(days)} & \colhead{(\mpersec)} & \colhead{(\mpersec)} & \colhead{(\mpersec)} & \colhead{(\mpersec)} & \colhead{(\mpersec)} & \colhead{(\mpersec)} &(dex) &(dex) & (dex) & (dex) &(dex) &(dex) &(dex)&(dex)\\
}
\startdata
7735.617860 & -8.14 & 2.20 & -5.15 & 2.95 & -10.17 & 2.45 & 0.5947 & 0.0031 & 0.4396 & 0.0031 & 0.4164 & 0.0030 & 0.9516 & 0.0028 \\
7747.734170 & -8.86 & 2.74 & -9.81 & 2.67 & -8.15 & 2.27 & 0.5971 & 0.0026 & 0.4450 & 0.0027 & 0.4145 & 0.0025 & 0.9548 & 0.0026 \\
7752.685530 & -5.28 & 1.76 & -7.02 & 2.14 & -3.99 & 1.87 & 0.6018 & 0.0023 & 0.4470 & 0.0022 & 0.4183 & 0.0021 & 0.9533 & 0.0022 \\
7755.711910 & -5.63 & 2.05 & -3.51 & 2.64 & -7.15 & 2.24 & 0.6042 & 0.0027 & 0.4569 & 0.0027 & 0.4237 & 0.0025 & 0.9683 & 0.0026 \\
7759.696560 & -9.93 & 2.58 & -8.25 & 3.36 & -11.08 & 2.70 & 0.5938 & 0.0032 & 0.4388 & 0.0033 & 0.4206 & 0.0031 & 0.9528 & 0.0034 \\
7762.686550 & -7.33 & 2.28 & -6.37 & 2.30 & -8.08 & 1.98 & 0.5855 & 0.0024 & 0.4434 & 0.0023 & 0.4075 & 0.0022 & 0.9539 & 0.0023 \\
7766.737730 & -13.49 & 3.27 & -12.96 & 3.95 & -13.82 & 3.12 & 0.5927 & 0.0037 & 0.4325 & 0.0038 & 0.4112 & 0.0036 & 0.9500 & 0.0037 \\
7779.501760 & 1.97 & 2.91 & -1.12 & 4.15 & 3.97 & 3.34 & 0.5951 & 0.0043 & 0.4524 & 0.0044 & 0.4183 & 0.0041 & 0.9566 & 0.0041 \\
7787.481300 & -3.23 & 9.66 & -30.21 & 12.65 & 9.81 & 8.79 & 0.5837 & 0.0105 & 0.4613 & 0.0127 & 0.4138 & 0.0114 & 0.9671 & 0.0116 \\
7791.467500 & -8.05 & 4.83 & -15.53 & 6.46 & -3.63 & 4.95 & 0.5900 & 0.0062 & 0.4468 & 0.0068 & 0.4148 & 0.0064 & 0.9705 & 0.0065 \\
7794.611520 & -1.00 & 2.32 & -4.83 & 2.95 & 1.63 & 2.47 & 0.5982 & 0.0029 & 0.4428 & 0.0030 & 0.4037 & 0.0028 & 0.9401 & 0.0029 \\
7798.500510 & -4.63 & 3.19 & -6.88 & 4.10 & -3.19 & 3.28 & 0.5869 & 0.0039 & 0.4387 & 0.0042 & 0.4104 & 0.0039 & 0.9434 & 0.0040 \\
7806.509160 & 0.33 & 5.05 & -2.15 & 7.19 & 1.73 & 5.41 & 0.6069 & 0.0063 & 0.4423 & 0.0069 & 0.4151 & 0.0066 & 0.9587 & 0.0068 \\
7814.550200 & 0.32 & 1.93 & 1.40 & 2.16 & -0.47 & 1.89 & 0.5954 & 0.0023 & 0.4514 & 0.0022 & 0.4191 & 0.0021 & 0.9514 & 0.0021 \\
7817.513200 & 9.82 & 3.43 & 10.42 & 5.75 & 9.50 & 4.36 & 0.5906 & 0.0051 & 0.4520 & 0.0056 & 0.4036 & 0.0053 & 0.9479 & 0.0054 \\
7821.529830 & -3.86 & 1.64 & -5.67 & 2.08 & -2.43 & 1.84 & 0.6006 & 0.0022 & 0.4458 & 0.0022 & 0.4123 & 0.0021 & 0.9458 & 0.0021 \\
7828.484510 & -7.52 & 4.59 & -14.18 & 6.96 & -3.86 & 5.16 & 0.5960 & 0.0061 & 0.4551 & 0.0068 & 0.4004 & 0.0063 & 0.9473 & 0.0066 \\
7832.533410 & -10.36 & 2.22 & -12.34 & 2.23 & -8.90 & 1.94 & 0.5962 & 0.0023 & 0.4395 & 0.0023 & 0.4107 & 0.0021 & 0.9439 & 0.0022 \\
7848.477660 & 1.28 & 2.33 & 4.47 & 2.51 & -0.88 & 2.08 & 0.5971 & 0.0024 & 0.4410 & 0.0025 & 0.4074 & 0.0023 & 0.9520 & 0.0024 \\
7855.492080 & -0.39 & 1.83 & 2.47 & 2.80 & -2.28 & 2.27 & 0.5971 & 0.0027 & 0.4600 & 0.0028 & 0.4211 & 0.0026 & 0.9515 & 0.0026 \\
7856.441020 & 0.75 & 2.28 & -1.34 & 2.72 & 2.07 & 2.18 & 0.5974 & 0.0025 & 0.4568 & 0.0026 & 0.4226 & 0.0024 & 0.9469 & 0.0025 \\
7857.414140 & -2.27 & 2.10 & -3.74 & 2.70 & -1.26 & 2.22 & 0.6009 & 0.0026 & 0.4624 & 0.0027 & 0.4236 & 0.0025 & 0.9606 & 0.0025 \\
7858.429730 & -1.15 & 1.72 & -2.99 & 2.94 & -0.02 & 2.30 & 0.5974 & 0.0027 & 0.4533 & 0.0028 & 0.4181 & 0.0026 & 0.9424 & 0.0027 \\
7859.444190 & -7.37 & 2.41 & -9.47 & 3.53 & -6.08 & 2.77 & 0.5951 & 0.0032 & 0.4574 & 0.0034 & 0.4222 & 0.0032 & 0.9593 & 0.0032 \\
7860.428910 & -7.92 & 6.65 & -25.00 & 10.06 & 0.76 & 7.17 & 0.5864 & 0.0075 & 0.4677 & 0.0090 & 0.4082 & 0.0083 & 0.9472 & 0.0080 \\
7861.419190 & -7.22 & 1.92 & -8.54 & 2.31 & -6.30 & 1.93 & 0.5866 & 0.0023 & 0.4477 & 0.0023 & 0.4175 & 0.0022 & 0.9486 & 0.0022 \\
7862.453700 & -5.89 & 2.14 & -8.13 & 2.95 & -4.41 & 2.40 & 0.5962 & 0.0027 & 0.4510 & 0.0028 & 0.4100 & 0.0026 & 0.9562 & 0.0028 \\
7863.426410 & 9.66 & 11.91 & 4.14 & 15.66 & 12.35 & 10.86 & 0.5670 & 0.0103 & 0.4228 & 0.0132 & 0.4133 & 0.0123 & 0.9659 & 0.0121 \\
7864.480200 & 6.13 & 6.58 & 16.88 & 9.80 & 1.15 & 6.66 & 0.5847 & 0.0070 & 0.4553 & 0.0082 & 0.4243 & 0.0076 & 0.9681 & 0.0079 \\
7875.429690 & -12.79 & 4.59 & -10.20 & 6.83 & -14.18 & 5.02 & 0.5694 & 0.0055 & 0.4342 & 0.0062 & 0.4046 & 0.0058 & 0.9382 & 0.0058 \\
7876.398880 & -4.16 & 2.00 & -5.11 & 2.49 & -3.47 & 2.04 & 0.5891 & 0.0024 & 0.4370 & 0.0024 & 0.4073 & 0.0023 & 0.9467 & 0.0023 \\
7877.374190 & -7.56 & 2.13 & -8.94 & 2.81 & -6.69 & 2.23 & 0.5838 & 0.0026 & 0.4381 & 0.0027 & 0.4081 & 0.0026 & 0.9454 & 0.0026 \\
7881.362850 & 4.00 & 3.02 & 2.38 & 2.61 & 5.10 & 2.16 & 0.5904 & 0.0024 & 0.4390 & 0.0025 & 0.4070 & 0.0023 & 0.9488 & 0.0024 \\
7882.390120 & 4.34 & 1.89 & -0.42 & 2.80 & 7.28 & 2.21 & 0.5883 & 0.0026 & 0.4462 & 0.0027 & 0.4126 & 0.0025 & 0.9412 & 0.0026 \\
7883.401660 & 0.53 & 4.48 & -2.95 & 7.00 & 2.36 & 5.06 & 0.5759 & 0.0056 & 0.4468 & 0.0063 & 0.4162 & 0.0058 & 0.9447 & 0.0057 \\
7886.415260 & -0.65 & 5.74 & -16.16 & 7.95 & 6.84 & 5.52 & 0.5926 & 0.0059 & 0.4490 & 0.0068 & 0.4333 & 0.0064 & 0.9300 & 0.0062 \\
7887.447050 & -0.34 & 1.85 & -1.02 & 2.38 & 0.13 & 1.89 & 0.5991 & 0.0023 & 0.4534 & 0.0024 & 0.4152 & 0.0022 & 0.9499 & 0.0023 \\
7888.414710 & -1.54 & 1.76 & -3.15 & 2.51 & -0.51 & 2.00 & 0.5978 & 0.0024 & 0.4468 & 0.0025 & 0.4164 & 0.0023 & 0.9451 & 0.0024 \\
7889.433290 & -1.13 & 2.41 & -2.06 & 3.27 & -0.53 & 2.58 & 0.5985 & 0.0032 & 0.4580 & 0.0034 & 0.4233 & 0.0032 & 0.9600 & 0.0032 \\
7890.451800 & 0.05 & 2.04 & 1.14 & 2.58 & -0.62 & 2.05 & 0.6074 & 0.0026 & 0.4651 & 0.0027 & 0.4219 & 0.0025 & 0.9715 & 0.0026 \\
7891.373770 & 3.25 & 2.53 & 4.26 & 2.48 & 2.98 & 1.99 & 0.5969 & 0.0025 & 0.4616 & 0.0026 & 0.4234 & 0.0024 & 0.9538 & 0.0024 \\
7892.398400 & 4.59 & 3.42 & -3.05 & 4.61 & 8.60 & 3.36 & 0.5972 & 0.0040 & 0.4538 & 0.0044 & 0.4246 & 0.0041 & 0.9582 & 0.0042 \\
7893.377370 & -2.67 & 3.33 & -6.62 & 4.96 & -0.59 & 3.59 & 0.6067 & 0.0044 & 0.4740 & 0.0049 & 0.4449 & 0.0047 & 0.9609 & 0.0049 \\
7894.381700 & -7.52 & 2.09 & -9.92 & 2.42 & -6.01 & 1.98 & 0.6021 & 0.0025 & 0.4630 & 0.0027 & 0.4263 & 0.0025 & 0.9576 & 0.0025 \\
7896.370090 & -5.15 & 2.17 & -4.23 & 2.38 & -5.61 & 1.93 & 0.6054 & 0.0025 & 0.4594 & 0.0026 & 0.4289 & 0.0024 & 0.9624 & 0.0024 \\
7897.357930 & -8.27 & 2.87 & -7.76 & 2.58 & -8.60 & 2.12 & 0.6059 & 0.0026 & 0.4591 & 0.0027 & 0.4195 & 0.0024 & 0.9571 & 0.0025 \\
7898.391440 & -7.20 & 2.30 & -3.54 & 2.66 & -9.76 & 2.24 & 0.6047 & 0.0027 & 0.4663 & 0.0028 & 0.4258 & 0.0026 & 0.9666 & 0.0027 \\
7901.415500 & -1.43 & 2.83 & -1.90 & 3.36 & -1.28 & 2.56 & 0.5935 & 0.0032 & 0.4546 & 0.0034 & 0.4245 & 0.0032 & 0.9539 & 0.0033 \\
7905.431390 & -9.57 & 2.48 & -10.72 & 4.20 & -8.89 & 3.28 & 0.6014 & 0.0041 & 0.4462 & 0.0044 & 0.4122 & 0.0042 & 0.9507 & 0.0041 \\
7909.422030 & -2.29 & 3.36 & -0.58 & 3.85 & -3.36 & 2.85 & 0.5819 & 0.0034 & 0.4454 & 0.0037 & 0.4080 & 0.0034 & 0.9429 & 0.0036 \\
7911.388840 & -2.44 & 1.61 & -0.28 & 2.08 & -3.84 & 1.69 & 0.5904 & 0.0021 & 0.4372 & 0.0022 & 0.4024 & 0.0020 & 0.9472 & 0.0021 \\
7912.363270 & -0.17 & 2.34 & 0.24 & 2.34 & -0.40 & 1.86 & 0.5940 & 0.0023 & 0.4392 & 0.0023 & 0.4070 & 0.0022 & 0.9493 & 0.0022 \\
7915.399610 & -3.79 & 2.95 & -2.82 & 4.28 & -4.43 & 3.43 & 0.5838 & 0.0039 & 0.4401 & 0.0043 & 0.4199 & 0.0040 & 0.9374 & 0.0039 \\
7916.378800 & -0.18 & 2.74 & -2.60 & 3.11 & 1.45 & 2.52 & 0.5918 & 0.0029 & 0.4366 & 0.0030 & 0.4119 & 0.0028 & 0.9379 & 0.0029 \\
7918.393180 & 4.62 & 2.17 & 4.93 & 2.63 & 4.34 & 2.26 & 0.5940 & 0.0027 & 0.4444 & 0.0028 & 0.4109 & 0.0026 & 0.9468 & 0.0026 \\
7919.379830 & 4.09 & 2.43 & 6.14 & 2.62 & 2.58 & 2.31 & 0.5949 & 0.0027 & 0.4491 & 0.0028 & 0.4092 & 0.0026 & 0.9403 & 0.0025 \\
7921.378720 & -1.47 & 1.82 & -1.19 & 2.64 & -1.70 & 2.23 & 0.5922 & 0.0028 & 0.4508 & 0.0029 & 0.4154 & 0.0027 & 0.9469 & 0.0026 \\
7924.380800 & 0.59 & 2.67 & 0.00 & 2.67 & 0.90 & 2.25 & 0.6027 & 0.0026 & 0.4565 & 0.0027 & 0.4222 & 0.0025 & 0.9534 & 0.0026 \\
\enddata
\end{deluxetable*}
\end{longrotatetable}

\end{document}